\documentclass{ws-ijmpd}
\usepackage[super,compress]{cite}
\usepackage{upgreek}

\newcommand{\A}{{\scriptscriptstyle{A}}}
\newcommand{\B}{{\scriptscriptstyle{B}}}
\newcommand{\C}{{\scriptscriptstyle{C}}}
\newcommand{\D}{{\scriptscriptstyle{D}}}
\newcommand{\I}{{\scriptscriptstyle{I}}}
\newcommand{\J}{{\scriptscriptstyle{J}}}
\newcommand{\M}{{\scriptscriptstyle{M}}}
\newcommand{\N}{{\scriptscriptstyle{N}}}
\newcommand{\K}{{\scriptscriptstyle{K}}}

\begin{document}

\newcounter{count}

\markboth{William G. Cook {\em et al.}}
{Dimensional reduction in numerical relativity: Modified cartoon formalism and regularization}

\catchline{}{}{}{}{}
%
%%%%%%%%%%%%%%%%%%%%%%%%%%%%%%%%%%%%%%%%%%%%%%%%%%%%%%%%%%%%%%%%%%%%

\title{Dimensional reduction in numerical relativity: Modified cartoon formalism and regularization
}

\author{WILLIAM G. COOK${}^1$, PAU FIGUERAS${}^{2,1}$, MARKUS KUNESCH${}^1$,
        ULRICH SPERHAKE${}^{1,3,4}$, SARAN TUNYASUVUNAKOOL${}^{2,1}$}

\address{${}^1\,$Department of Applied Mathematics and Theoretical Physics \\
University of Cambridge, Cambridge CB3 0WA, United Kingdom
%wc259@cam.ac.uk
\\[5pt]
${}^2\,$School of Mathematical Sciences\\
Queen Mary University of London,
Mile End Road, London E1 4NS, United Kingdom
%p.figueras@qmul.ac.uk
\\[5pt]
${}^3\,$California Institute of Technology, Pasadena, CA 91125, USA \\[5pt]
%U.Sperhake@damtp.cam.ac.uk
${}^4\,$Department of Physics and Astronomy, The University of Mississippi,
University, MS 38677, USA\\ [5pt]
}

\maketitle

\begin{history}
\received{--- 2016}
\revised{--- 2016}
\end{history}

\begin{abstract}
We present in detail the Einstein equations
in the Baumgarte-Shapiro-Shibata-Nakamura formulation
for the case of $D$ dimensional spacetimes with $SO(D-d)$
isometry based on a method originally introduced
in \cite{Pretorius:2004jg}. Regularized
expressions are given for a numerical implementation of this
method on a vertex centered grid including the origin
of the quasi-radial coordinate that covers the extra dimensions
with rotational symmetry. Axisymmetry, corresponding to the
value $d=D-2$, represents a special case with fewer constraints
on the vanishing of tensor components and is conveniently
implemented in a variation of the general method. The robustness
of the scheme is demonstrated for the case of a black-hole
head-on collision in $D=7$ spacetime dimensions with $SO(4)$ symmetry.
\end{abstract}


\ccode{PACS numbers: 04.25.D-, 04.70.-s, 04.50.-h, 04.25.dg}
% 04.25.D-: Numerical relativity 
% 04.70.-s: Physics of black holes
% 04.50.-h: Gravity in more than four dimensions, Kaluza-Klein theory,
%           unified field theories; alternative theories of gravity
% 04.25.dg: Numerical studies of black holes and black-hole binaries

%\tableofcontents

\section{Introduction}
For most of its history, numerical relativity,
i.e.,~the construction of solutions to Einstein's
field equations with numerical methods,
was mainly motivated
by the modeling of compact objects as sources of gravitational
waves\cite{Abbott:2016blz}
for ground based [Laser Interferometer Gravitational-Wave
Observatory (LIGO), Virgo] and space based
[Laser Interferometer Space Antenna (LISA)]
detectors; see for
example \cite{Read:2013zra,Hinder:2013oqa}. Following
the breakthroughs in black-hole (BH) binary simulations
in 2005 \cite{Pretorius:2005gq,Baker:2005vv,Campanelli:2005dd},
however, the field rapidly expanded into a variety of new physics
frontiers \cite{Cardoso:2014uka}.

These applications often involve higher-dimensional
spacetimes where BHs are known from (semi-)analytic studies
to exhibit a richer phenomenology as for example through the
existence of topologically
non-spherical horizons or gravitational instabilities
\cite{Emparan:2008eg}. Despite
considerable progress through analytic, perturbative and numerical
methods, our understanding of the properties of these higher-dimensional
BHs is still a long way from the level of maturity
obtained in the four-dimensional case. Yet, applications of numerical relativity
to BHs in $D>4$ have already revealed a plethora of exciting results.

Critical spin parameters have been identified above which
Myers-Perry\footnote{These BHs are the higher dimensional analogues of the Kerr solution \cite{Myers:1986un}.} BHs become unstable to bar mode perturbations
and migrate to more slowly spinning BHs via GW emission.
\cite{Shibata:2009ad,Shibata:2010wz}.
The celebrated Gregory-Laflamme instability\cite{Gregory:1993vy} has been shown to lead to the formation
of naked singularities in finite asymptotic time in numerical simulations of black strings in $D=5$ dimensions.
\cite{Lehner:2011wc} Most recently, a similar behaviour
has been identified in evolutions of thin black rings
demonstrating the first violation of cosmic censorship for
a generic type of asymptotically flat initial data
\cite{Figueras:2015hkb}; see also \cite{Santos:2015iua} for
a perturbative study.
Applications of the gauge-gravity duality
often consider $D=5$ dimensional BHs in asymptotically Anti de Sitter (AdS)
spacetimes such that the dual
Conformal Field Theory (CFT) lives on the $D=4$ dimensional
(conformal) boundary of the spacetime. Applications of this AdS/CFT correspondence
include the thermalization of quark-gluon plasma, turbulence or
jet quenching in heavy-ion collisions; see
\cite{Bantilan:2012vu,Chesler:2013lia,Bantilan:2014sra,
Gubser:2014qua,Chesler:2015bba,Buchel:2015saa} and references therein.

BH collisions provide fertile ground for numerical relativity in higher
dimensions. First, we obtain unprecedented insight into the
dynamics of general relativity in its most violent, non-linear
regime. Furthermore, the so-called TeV
gravity scenarios provide solutions to the hierarchy problem
in terms of large ($\sim$ sub millimeter) extra dimensions
\cite{Antoniadis:1990ew,Antoniadis:1998ig,ArkaniHamed:1998rs}
or infinite extra dimensions with a warp factor
\cite{Randall:1999ee,Randall:1999vf} which are accessible to gravity
but no other standard-model interactions. If correct, these
theories open up the possibility of BH formation in
particle collisions or in cosmic ray showers
\cite{Banks:1999gd,Dimopoulos:2001hw,Giddings:2001bu}.
Valuable input for the analysis of experimental data at the
Large Hadron Collider includes the scattering cross-section and
energy loss in GWs. Numerical relativity has provided us with
a rather comprehensive understanding of these collisions
in $D=4$ \cite{Sperhake:2008ga,Shibata:2008rq,Choptuik:2009ww,
Sperhake:2009jz,
East:2012mb,Rezzolla:2012nr,Sperhake:2012me,Healy:2015mla,Sperhake:2015siy},
which the community now starts extending to higher $D$
\cite{Witek:2010xi,Witek:2010az,Okawa:2011fv,Witek:2014mha}.
For more details on these new areas in numerical relativity research
see \cite{Cardoso:2014uka}.

Numerical simulations of BH spacetimes in higher dimensions
are a challenging task. First and foremost this is simply a
consequence of the required computational resources. Simulations
in $D=4$ require of the order of $\mathcal{O}(10^2)$ cores
and $\mathcal{O}(10^2)~{\rm Gb}$ of memory. Each extra spatial
dimension introduces an additional factor of $\mathcal{O}(10^2)$
grid points and correspondingly more memory and floating point
operations. Even with modern high-performance computing systems,
this sets practical limits on the feasibility of accurately
evolving higher-dimensional spacetimes. At the same time,
many of the outstanding questions can be addressed by imposing
symmetry assumptions on the spacetimes in question such as
planar symmetry in modeling asymptotically AdS spacetimes
\cite{Chesler:2013lia}, cylindrical symmetry for black
strings \cite{Lehner:2011wc} or different types of rotational
symmetries \cite{Figueras:2015hkb}. This can be achieved in
practice by either (i) using a specific form of the line element
that directly imposes the symmetry in question
(see e.g.~\cite{Chesler:2013lia}), (ii) starting with a generic
line element and applying dimensional reduction through
isometry (see e.g.~\cite{Geroch:1970nt,Zilhao:2010sr,Zilhao:2013gu})
or (iii) implementing the symmetry through a so-called
{\em Cartoon method} \cite{Alcubierre:1999ab}. Here we are
concerned with the latter approach and, more specifically, with
a modification thereof originally introduced in
\cite{Pretorius:2004jg} (see also
\cite{Shibata:2010wz,Yoshino:2011zz,Yoshino:2011zza})
which we will henceforth refer to as
the {\em modified Cartoon method}.

This article is structured as follows. In Sec.~\ref{sec:cartoon},
we introduce the notation used throughout our work,
and illustrate the modified Cartoon implementation of the symmetries for a specific
example. In Sec.~\ref{sec:BSSNdelta} we introduce the
Baumgarte-Shapiro-Shibata-Nakamura\cite{Shibata:1995we,Baumgarte:1998te}
(BSSN) evolution system we use for the Einstein equations,
and derive their specific form
in $SO(D-d)$ symmetry when rotational symmetry is present in
$\ge 2$ planes which corresponds to $d < D-2$. The axisymmetric
case $d=D-2$ imposes less restrictive conditions on the vanishing
of tensor density components and their derivatives and the
particulars of its numerical implementation are discussed in
Sec.~\ref{sec:SO2}. As an example, we present
in Sec.~\ref{sec:application} numerical simulations
of a BH head-on collision in $D=7$ dimensions employing $SO(4)$ symmetry.
We summarize our findings in Sec.~\ref{sec:conclusions}
and include in three appendices
a list of important relations for the components
of tensors and derived quantities as well as the
regularization necessary at the origin in the quasi-radial direction.

%=============================================================================
\section{$SO(D-d)$ symmetry in the modified Cartoon method}
\label{sec:cartoon}
%
%======================================
\subsection{Coordinates}
It is instructive to illustrate the method by considering first a
simpler scenario: axisymmetry in three spatial dimensions. Let $(x,\,z,\,w)$
denote Cartesian coordinates and assume rotational symmetry about
the $x$ axis\footnote{It is more common to label the coordinates
$(x,\,y,\,z)$ and use symmetry about the $z$ axis, but our choice
of labels emphasizes
more clearly the analogy to the higher-dimensional case.}
i.e.,~there exists a rotational Killing field in the $z,\,w$ plane.
Evidently, the geometry of such a three-dimensional manifold
can be constructed straightforwardly provided all tensors (e.g.~the
metric) are known on the semi infinite plane
$w=0$, $z\ge 0$, $x\in \mathbb{R}$. We note the simplification in the
computational task: the $w$ coordinate has dropped out and the
quasi-radius $z$ takes on only non-negative values, reducing an
originally three-dimensional computational domain to a calculation
on half of $\mathbb{R}^2$. This is the case
considered in the original papers \cite{Alcubierre:1999ab,Pretorius:2004jg}.

The most common applications will likely
consider higher-dimensional spacetimes with
$SO(D-3)$ symmetry,
but here we present the general application to a $D$ dimensional spacetime with $SO(D-d)$ symmetry, where
$d \in \mathbb{N},~1\le d \le D-2$. Let us then consider
a $D$ dimensional spacetime
consisting of a manifold $\mathcal{M}$ and a metric $g_{\A\B}$
of signature $D-2$ where $A,\,B,\,\ldots = 0,\,\ldots,\,D-1$. We assume
the spacetime to obey $SO(D-d)$ symmetry and introduce Cartesian
coordinates by
\begin{equation}
  X^{\A}=(t,\,\underbrace{x^1,\,x^2,\ \ldots \ x^{d-1}}_{(d-1)\times},
        \,z,\,\underbrace{w^{d+1},\,w^{d+2},\,\ldots,\,
                          w^{D-1}}_{(D-d-1)\times}) =:
        (t,\,x^{\hat{i}},\,z,\,w^a)\,,
\end{equation}
where $\hat{i}=1,\ldots d-1,~~a=d+1,\,\ldots,\,D-1$. $SO(D-d)$ symmetry implies the existence of
rotational Killing vectors in each plane spanned by two of the
coordinates $(z,\,w^a)$. In complete analogy with the axisymmetric
scenario discussed above, it
is now sufficient to provide data on the $d$-dimensional semi-infinite
hyperplane $w^a=0, x^{\hat{i}}\in \mathbb{R}$, $z\ge 0$. The components of a tensor
at any point in the spacetime can then be obtained by appropriately
rotating data from the hyperplane onto the point in question.

In modeling spacetimes with such symmetries, it is therefore
entirely sufficient to compute data on the hyperplane which
largely solves the problem of increased computational cost
mentioned in the introduction. There remains, however, the difficulty
that the Einstein equations, irrespective of the specific formulation
one chooses, contain derivatives of tensor components in the
$w^a$ directions which cannot be evaluated numerically in the usual
fashion, as for example using finite differences or spectral methods.
Furthermore, the number of tensor components present in the Einstein
equations still increases rapidly with the dimension parameter $D$
resulting in a substantial increase of memory requirements and
floating point operations. Both of these difficulties are overcome
by exploiting the conditions imposed on the tensor components and
their derivatives by the $SO(D-d)$ symmetry.
It is these conditions which we address next. It turns out to be
convenient in this discussion to distinguish
between (1) the case $d=D-2$ corresponding
to $SO(2)$ isometry, and (2) all remaining cases $d<D-2$.
We defer discussion of the special case $d=D-2$
to Sec.~\ref{sec:SO2} where we present a numerical treatment
specifically designed for conveniently dealing with it.
The description of this treatment will be simpler
after first handling the class of symmetries with
$d<D-2$ which we discuss in the remainder of this
and in the next section.

%======================================
\subsection{Tensor components in $SO(D-d)$
symmetry for $d<D-2$}
\label{sec:components}
The key ingredient we use in reducing the number of independent
tensor components and relating their derivatives are the rotational
Killing vectors and the use of coordinates adapted to the integral
curves of these Killing vectors. The method is best introduced by
considering a concrete example. Let $\boldsymbol{\xi}$ denote
the Killing vector field corresponding to the rotational symmetry
in the $(z,w)$ plane, where $w\equiv w^a$ for some fixed number
$a\in\{d+1,\,\ldots,\,D-1\}$. We introduce a new coordinate system
that replaces $(z,w)$ with cylindrical coordinates and leaves
all other coordinates unchanged,
\begin{eqnarray}
  && \bar{X}^A = (t,\,x^{\hat{i}},\,\rho,\,w^{d+1},\,\ldots,\,w^{a-1},\,\varphi,\,
        w^{a+1},\,\ldots,\,w^{D-1})\,, \\
  && \rho=\sqrt{z^2+w^2}\,,~~~~~~~~~~\,~~~z=\rho \cos \varphi\,, \\
  && \varphi = {\rm arctan} \frac{w}{z}\,,~~~~~~~~~~~~~~~
        w = \rho \sin \varphi\,.
\end{eqnarray}
In these coordinates, the Killing field is $\boldsymbol{\xi}
= \boldsymbol{\partial}_{\varphi}$ and the vanishing of the
Lie derivative $\mathcal{L}_{\boldsymbol{\xi}}g_{\A\B}=0$
implies $\partial_{\varphi}g_{\A\B}=0$. Note that quantities constructed
from the spacetime metric directly inherit this property. This
applies, in particular, to the Arnowitt-Deser-Misner\cite{Arnowitt:1962hi}
(ADM) -- see also \cite{York1979,Gourgoulhon:2007ue} --
and the BSSN variables widely
used in numerical relativity. For $d<D-2$, one can furthermore
show that the $\varphi$ component of a vector field
and those components of a tensor field $T_{\A\B}$, where exactly
one index is $\varphi$, vanish\footnote{Here the
case $d=D-2$ represents an exception; an axisymmetric, toroidal
magnetic field, for example, satisfies $SO(2)$ symmetry, but has
a non-vanishing $\varphi$ component.}.

The concrete example we now discuss in more detail concerns a symmetric
tensor density $T_{\A\B}$ of weight $\mathcal{W}$ and, in particular,
the mixed components $T_{iw}$, where the index $i$ stands for any one of
the $(x^{\hat{i}},z)$ coordinates
and $w$ stands for one of the $w^a$.
We first consider the components $T_{\hat{i}w}$
for some fixed value of $\hat{i}$. Transforming the component $\bar T_{\hat i \varphi}$ to Cartesian coordinates, one gets
\begin{equation}
  \bar{T}_{\hat{i}\varphi} = \mathcal{D}^{\mathcal{W}}
        \frac{\partial X^{\A}}{\partial \bar{X}^{\hat{i}}}
        \frac{\partial X^{\B}}{\partial \varphi}
        T_{\A\B}\,
        = \mathcal{D}^{\mathcal{W}} \left(
        - w\,T_{\hat{i} z} + z\, T_{\hat{i}w}
        \right)\,,
\end{equation}
where $\mathcal{D}$ is the Jacobian $\det (\partial X^{\A}/\partial
\bar{X}^{\B}) =\rho$. Using that $\bar{T}_{\hat{i}\varphi}=0$ by symmetry, this equation implies
\begin{equation}
  T_{\hat{i}w} = \frac{w}{z}\,T_{\hat{i}z}\,. \label{eq:Txyw}
\end{equation}
Similarly, transforming $\bar{T}_{\rho\varphi}$ to Cartesian coordinates and using that $\bar{T}_{\rho\varphi}=0$ by symmetry, one straightforwardly gets
\begin{equation}
  T_{zw} = \frac{zw}{z^2-w^2}(T_{zz}-T_{ww})\,. \label{eq:Tzw}
\end{equation}
Recalling that the computational domain is the hyperplane
$w^a=0$, $x^{\hat{i}}\in \mathbb{R}$, $z\ge 0$, we conclude from
Eqs.~(\ref{eq:Txyw}),(\ref{eq:Tzw}) that on the computational domain
$T_{iw}=0$. This argument holds for any specific choice of the
coordinate $w$, so that we conclude
\begin{equation}
  T_{ia} = 0\,.
\end{equation}
To compute the derivatives with respect to $w$ on the $w=0$ hyperplane, one can proceed as follows. For the tensor components in the example above, one can simply use \eqref{eq:Txyw} and \eqref{eq:Tzw} to calculate $\partial_w T_{ia}$ and then set $w=0$. Alternatively, writing the Killing field $\boldsymbol{\xi}$ as
\begin{equation}
  \boldsymbol{\xi} = z\,\boldsymbol{\partial_w}-w\,\boldsymbol{\partial_z}\,,
\end{equation}
and imposing the vanishing of the Lie derivative $\mathcal{L}_{\boldsymbol{\xi}}T_{ia}=0$ on the $w=0$ hyperplane, one gets
\begin{equation}
  \partial_w T_{iw} = \frac{T_{iz} - \delta_{iz}T_{ww}}{z}\,.
\end{equation}
Repeating this process for all components of scalar, vector and
rank 2 tensor densities as well as their first and second derivatives,
we get the relations summarized in \ref{app:cartoon}.

We have shown the calculation here explicitly for the case of
tensor densities.
It can be shown that the vectorial expressions thus obtained also apply to
the contracted Christoffel symbol
$\Gamma^{\A}\equiv g^{\M\N} \Gamma^{\A}_{\M\N}$ constructed from the
metric, even though it is not a vector density.
%=============================================================================

\section{Dimensional reduction of the BSSN equations}
\label{sec:BSSNdelta}
In this section, we will apply the symmetry relations obtained above
to the specific case of the BSSN formulation of the Einstein
equations in $D$ spacetime dimensions. We emphasize, however, that
the procedure spelled out here for the BSSN system can be applied
in similar form to any of the alternative popular formulations used
in numerical relativity.

%======================================
\subsection{The $D$ dimensional BSSN equations}
The starting point for the BSSN formulation is a space-time, or
$(D-1)+1$, split where the spacetime is foliated in terms of
a one-parameter family of $D-1$ dimensional, spatial hypersurfaces.
In coordinates adapted to this split, the line element takes on the form
\begin{equation}
  ds^2 = g_{\A\B}dx^{\A} dx^{\B}
        = (\alpha^2 + \beta_{\I} \beta^{\I})dt^2
        +2\beta_{\I} dx^{\I} dt + \gamma_{\I\J} dx^{\I} dx^{\J}\,,
\end{equation}
where $I,\,J,\,\ldots = 1,\,\ldots,\,D-1$ and $\alpha$ and $\beta^{\I}$
denote the lapse function and shift vector, respectively. The ADM
equations in the form developed by York\cite{York1979} then
result in one Hamiltonian constraint, $D-1$ momentum constraints
and $D(D-1)/2$ second-order evolution equations for the
spatial metric components $\gamma_{\I\J}$. The latter are formulated as
a first-order-in-time system by introducing the extrinsic curvature
$K_{\I\J}$ through
\begin{equation}
  \partial_t \gamma_{\I\J} = \beta^{\M}\partial_{\M} \gamma_{\I\J}
        + \gamma_{\M\J}\partial_{\I} \beta^{\M}
        + \gamma_{\I\M}\partial_{\J} \beta^{\M}
        -2\alpha K_{\I\J}\,.
\end{equation}
Space-time decompositions of the Einstein equations typically
split the energy momentum tensor analogously into time, space
and mixed components according to
\begin{equation}
  \rho \equiv T_{\A \B} n^{\A} n^{\B}\,,~~~
        j_{\A} \equiv -(\delta^{\B}{}_{\A} + n^{\B} n_{\A})T_{\B\C}n^{\C}\,,~~~
        S_{\A \B} \equiv (\delta^{\C}{}_{\A}+n^{\C}n_{\A})
                (\delta^{\D}{}_{\B}+n^{\D}n_{\B}) T_{\C\D}\,,
\end{equation}
where $n^{\A}$ denotes the future pointing, timelike unit normal field on
the spatial hypersurfaces.
The complete set of the ADM equations, thus obtained, can be found as
Eqs.~(52)-(55) in \cite{Cardoso:2014uka}.

The BSSN system is obtained
from the ADM equations by applying a conformal transformation to the
spatial metric, a trace split of the extrinsic curvature and
promotion of the contracted spatial Christoffel symbols to the status of
evolution variables. The BSSN variables are defined as
\begin{eqnarray}
  &\chi = \gamma^{-1/(D-1)}\,,
  ~~~~~
  &K = \gamma^{\M \N} K_{\M \N}, \nonumber \\
  &\tilde{\gamma}_{\I \J} = \chi \gamma_{\I \J}
  ~~~
  &\Leftrightarrow
  ~~~
  \tilde{\gamma}^{\I \J} = \frac{1}{\chi} \gamma^{\I \J}, \nonumber \\
  &\tilde{A}_{\I \J} = \chi \left( K_{\I \J} - {\displaystyle \frac{1}{D-1}}
      \gamma_{\I \J} K \right)
  ~~~
  &\Leftrightarrow
  ~~~
  K_{\I \J} = \frac{1}{\chi} \left( \tilde{A}_{\I \J} + \frac{1}{D-1}
      \tilde{\gamma}_{\I \J} K \right)\,, \nonumber \\[10pt]
  &\tilde{\Upgamma}^{\I} = \tilde{\gamma}^{\M \N}
        \tilde{\Upgamma}^{\I}_{\M \N}\,,&
  \label{eq:BSSNvars}
\end{eqnarray}
where $\gamma = \det \gamma_{IJ}$, and $\tilde{\Upgamma}^{\I}_{\M\N}$
are the Christoffel symbols associated with the conformal metric
$\tilde{\gamma}_{IJ}$. We formulate here the conformal factor in
terms of the variable $\chi$, following \cite{Campanelli:2005dd}.
Alternative versions of the equations using variables
$W\equiv \sqrt{\chi}$ or $\phi \equiv -(\ln \chi)/4$
can be found in \cite{Marronetti:2007wz,Alcubierre:2002kk}. Note that
the definition of the BSSN variables in (\ref{eq:BSSNvars}) implies
two algebraic and one differential constraints given by
\begin{equation}
  \tilde{\gamma} = 1,~~~~~\tilde{\gamma}^{\M \N} \tilde{A}_{\M \N} = 0,~~~~~
  \mathcal{G}^{\I} \equiv \tilde{\Upgamma}^{\I} - \tilde{\gamma}^{\M \N}
        \tilde{\Upgamma}^{\I}_{\M \N}=0\,.
\end{equation}
The $D$ dimensional BSSN equations are then given by the Hamiltonian and
momentum constraints
\begin{eqnarray}
  \mathcal{H} &\equiv& \mathcal{R} + \frac{D-2}{D-1}K^2
        - \tilde{A}^{\M\N}\tilde{A}_{\M\N} - 16\pi \rho - 2\Lambda = 0\,,
        \label{eq:BSSNHam} \\
  \mathcal{M}_{\I} &\equiv& \tilde{\gamma}^{\M\N}
        \tilde{D}_{\M} \tilde{A}_{\N\I} - \frac{D-2}{D-1}
        \partial_{\I} K - \frac{D-1}{2} \tilde{A}^{\M}{}_{\I}
        \frac{\partial_{\M} \chi}{\chi}
        - 8 \pi j_{\I} = 0\,,
        \label{eq:BSSNmom}
\end{eqnarray}
and the evolution system
\begin{eqnarray}
  \partial_t \chi &=& \beta^{\M} \partial_{\M} \chi
        + \frac{2}{D-1} \chi (\alpha K - \partial_{\M} \beta^{\M})\,,
        \label{eq:BSSNdtchi} \\
  \partial_t \tilde{\gamma}_{\I \J} &=& \beta^{\M} \partial_{\M}
        \tilde{\gamma}_{\I \J}
        + 2\tilde{\gamma}_{\M (\I} \partial_{\J )} \beta^{\M}
        - \frac{2}{D-1}\tilde{\gamma}_{\I \J} \partial_{\M} \beta^{\M}
        - 2\alpha \tilde{A}_{\I \J}\,,\\
  \partial_t K &=& \beta^{\M} \partial_{\M} K
        - \chi \tilde{\gamma}^{\M \N} D_{\M} D_{\N} \alpha
        + \alpha \tilde{A}^{\M \N}\tilde{A}_{\M \N}
        + \frac{1}{D-1}\alpha K^2 \nonumber \\
     && + \frac{8\pi}{D-2} \alpha [S+(D-3)\rho]
        - \frac{2}{D-2}\alpha \Lambda\,, \\
  \partial_t \tilde{A}_{\I \J} &=& \beta^{\M} \partial_{\M} \tilde{A}_{\I \J}
        + 2\tilde{A}_{\M (\I} \partial_{\J )} \beta^{\M}
        - \frac{2}{D-1} \tilde{A}_{\I \J} \partial_{\M} \beta^{\M}
        + \alpha K\tilde{A}_{\I \J}
        - 2\alpha \tilde{A}_{\I \M} \tilde{A}^{\M}{}_{\J} \nonumber \\
     && + \chi \left(
          \alpha \mathcal{R}_{\I \J} - D_{\I} D_{\J} \alpha
          - 8\pi \alpha S_{\I \J} \right)^{\rm TF}\,,\\
  \partial_t \tilde{\Upgamma}^{\I} &=& \beta^{\M} \partial_{\M}
          \tilde{\Upgamma}^{\I}
        + \frac{2}{D-1} \tilde{\Upgamma}^{\I} \partial_{\M} \beta^{\M}
        - \tilde{\Upgamma}^{\M}\partial_{\M} \beta^{\I}
        + \tilde{\gamma}^{\M \N} \partial_{\M} \partial_{\N} \beta^{\I}
        + \frac{D-3}{D-1}\tilde{\gamma}^{\I \M} \partial_{\M}
          \partial_{\N} \beta^{\N}
          \nonumber \\
     && - \tilde{A}^{\I \M} \left[
          (D-1) \alpha \frac{\partial_{\M} \chi}{\chi}
          + 2\partial_{\M} \alpha \right]
        + 2\alpha \tilde{\Upgamma}^{\I}_{\M \N} \tilde{A}^{\M \N}
        - 2\frac{D-2}{D-1} \alpha \tilde{\gamma}^{\I \M} \partial_{\M} K
        \nonumber \\
     && - 16\pi \frac{\alpha}{\chi} j^{\I}
        - \sigma \mathcal{G}^{\I} \partial_{\M}\beta^{\M}\,.
        \label{eq:BSSNdtGamma}
\end{eqnarray}
Here,
$\mathcal{R}_{\I\J}$, $\mathcal{R}$ are the Ricci tensor and scalar
associated with the physical spatial metric $\gamma_{\I\J}$,
$\Lambda$ is the cosmological constant,
the superscript ``TF'' denotes the trace-free part and we have added
a constraint damping term $\sigma \mathcal{G}^{\I}$ in the last line,
following the suggestion by \cite{Yo:2002bm}.
The above equations are complemented by the following
auxiliary relations,
\begin{eqnarray}
  \Upgamma^{\I}_{\J \K} &=& \tilde{\Upgamma}^{\I}_{\J \K}
        - \frac{1}{2\chi}\left( \delta^{\I}{}_{\K} \partial_{\J} \chi +
          \delta^{\I}{}_{\J} \partial_{\K} \chi
          - \tilde{\gamma}_{\J \K} \tilde{\gamma}^{\I \M} \partial_{\M} \chi
          \right)\,, \\
  \mathcal{R}_{\I \J} &=& \tilde{\mathcal{R}}_{\I \J}
        + \mathcal{R}^{\chi}_{\I \J}\,, \\
  \mathcal{R}^\chi_{\I \J} &=&
        \frac{\tilde{\gamma}_{\I \J}}{2\chi} \left[
        \tilde{\gamma}^{\M \N} \tilde{D}_\M \tilde{D}_\N \chi
        - \frac{D-1}{2\chi} \tilde{\gamma}^{\M \N}
          \partial_\M \chi \,\, \partial_\N \chi \right]
        \nonumber \\
     && + \frac{D-3}{2\chi} \left( \tilde{D}_\I \tilde{D}_\J \chi
          - \frac{1}{2\chi} \partial\I \chi\,\,\partial_\J \chi \right)\,,
        \\
  \tilde{\mathcal{R}}_{\I \J} &=&
        - \frac{1}{2} \tilde{\gamma}^{\M \N} \partial_{\M} \partial_{\N}
          \tilde{\gamma}_{\I \J}
        + \tilde{\gamma}_{\M(\I} \partial_{\J)} \tilde{\Upgamma}^{\M}
        + \tilde{\Upgamma}^{\M} \tilde{\Upgamma}_{(\I \J)\M}
        \nonumber \\
     && + \tilde{\gamma}^{\M \N} \left[
          2\tilde{\Upgamma}^{\K}_{\M(\I} \tilde{\Upgamma}_{\J)\K \N}
          + \tilde{\Upgamma}^{\K}_{\I \M} \tilde{\Upgamma}_{\K \J \N} \right]\,,
        \\
  D_{\I} D_{\J} \alpha &=&
        \tilde{D}_{\I} \tilde{D}_{\J} \alpha
        + \frac{1}{\chi}\partial_{(\I} \chi\,\partial_{\J )} \alpha
        - \frac{1}{2\chi}\tilde{\gamma}_{\I \J} \tilde{\gamma}^{\M \N}
          \partial_{\M} \chi\,\partial_{\N} \alpha\,.
        \label{eq:BSSNDDalpha}
\end{eqnarray}
The BSSN equations in this form are general and facilitate the numerical
construction of $D$ dimensional spacetimes. Next, we will describe in
detail how the equations can be reduced to an effective system
in $d$ spatial dimensions for spacetimes obeying rotational symmetry
with $d < D-2$.

%======================================
\subsection{The BSSN equations with $SO(D-d)$ symmetry for
$d<D-2$}
We now apply the relations summarized in \ref{app:cartoon}
to the definition of the BSSN variables (\ref{eq:BSSNvars})
and the
$D$ dimensional BSSN equations
(\ref{eq:BSSNdtchi})-(\ref{eq:BSSNdtGamma}). Recalling that early and
middle Latin indices run over
$a,\,b,\,\ldots=d+1,\,\ldots,\,D-1$ and
$i,\,j,\,\ldots=1,\,\ldots\,d$, respectively, and introducing $n\equiv D-d-1$,
the variables are given in terms of their ADM counterparts by
\begin{eqnarray}
    &\chi = \gamma^{-1/(D-1)},~~
    \gamma = \det \gamma_{\I\J}=\gamma_{ww}^n \det \gamma_{ij}\,,
  ~~~
  &K = \gamma^{\M \N}K_{\M\N}=\gamma^{mn}K_{mn} + n\gamma^{ww}K_{ww},
  \nonumber \\
  &\tilde{\gamma}_{ij} = \chi \gamma_{ij}\,,~~
  \tilde{\gamma}_{ww} = \chi \gamma_{ww}
  ~~~
  &\Leftrightarrow
  ~~~
  \tilde{\gamma}^{ij} = \frac{1}{\chi} \gamma^{ij}\,,~~
  \tilde{\gamma}^{ww} = \frac{1}{\chi} \gamma^{ww}\,, \nonumber \\
  &\tilde{A}_{ij} = \chi \left( K_{ij} - \frac{1}{D-1} \gamma_{ij} K
      \right)
  ~~~
  &\Leftrightarrow
  ~~~
  K_{ij} = \frac{1}{\chi}
      \left( \tilde{A}_{ij} + \frac{1}{D-1} \tilde{\gamma}_{ij}
      K \right)\,, \nonumber \\
  &\tilde{A}_{ww} = \chi \left( K_{ww} - \frac{1}{D-1} \gamma_{ww} K
      \right)
  ~~~
  &\Leftrightarrow
  ~~~
  K_{ww} = \frac{1}{\chi}
      \left( \tilde{A}_{ww} + \frac{1}{D-1} \tilde{\gamma}_{ww}
      K \right)\,, \nonumber \\
  &\tilde{\Upgamma}^i = \tilde{\gamma}^{\M\N} \tilde{\Upgamma}^i_{\M\N}
        = \tilde{\gamma}^{mn} \tilde{\Upgamma}^i_{mn}
          +n\tilde{\gamma}^{ww} \tilde{\Upgamma}^i_{ww}\,,
  \label{eq:BSSNvarsD}
\end{eqnarray}
where
\begin{equation}
  \tilde{\Upgamma}^i_{ww} = -\frac{1}{2} \tilde{\gamma}^{im}
        \partial_m \tilde{\gamma}_{ww}
        + \frac{\delta^i{}_z - \tilde{\gamma}^{zi} \tilde{\gamma}_{ww}}{z}\,.
        \label{eq:Gammaiww}
\end{equation}
We first note that the spatial metric with $SO(D-d)$ symmetry has the
form
\begin{equation}
  \tilde{\gamma}_{IJ} =
  \left(
    \begin{array}{cccc|cccc}
      \tilde{\gamma}_{x^1x^1} & \cdots & \tilde{\gamma}_{x^1x^{d-1}} & \tilde{\gamma}_{xz} & 0 & 0 & \cdots & 0 \\
      \vdots & \ddots &\vdots &\vdots & \vdots & \vdots & \cdots & \vdots \\
      \tilde{\gamma}_{x^{d-1}x^1} & \cdots & \tilde{\gamma}_{x^{d-1}x^{d-1}} & \tilde{\gamma}_{x^{d-1}z} & 0 & 0 & \cdots & 0 \\
      \tilde{\gamma}_{zx^1} & \cdots & \tilde{\gamma}_{zx^{d-1}} & \tilde{\gamma}_{zz} & 0 & 0 & \cdots & 0 \\
      \hline
      0   & \cdots   & 0      & 0      & \tilde{\gamma}_{ww} & 0 & \ldots & 0 \\
      0   & \cdots   & 0      & 0      & 0 & \tilde{\gamma}_{ww} & \ldots & 0 \\
      \vdots & \cdots & \vdots & \vdots & \vdots              & \vdots & \ddots & \vdots \\
      0      & \cdots & 0      & 0      & 0                   & 0 & \cdots & \tilde{\gamma}_{ww}
    \end{array}
  \right)
  \label{eq:gammamatrix}
  \,,
\end{equation}
which simplifies the calculation of the inverse metric $\tilde{\gamma}^{\A\B}$;
see \ref{app:regularization}.

The constraint equations (\ref{eq:BSSNHam}), (\ref{eq:BSSNmom}) become
\begin{eqnarray}
  \mathcal{H} &=& \chi \tilde{\gamma}^{mn} \mathcal{R}_{mn}
        - \tilde{A}^{mn} \tilde{A}_{mn}
        + \frac{D-2}{D-1} K^2
        + n \left( \chi \tilde{\gamma}^{ww} \mathcal{R}_{ww}
          - \frac{\tilde{A}_{ww}^2}{\tilde{\gamma}_{ww}^2} \right)
        \nonumber \\
     && -16\pi \rho - 2\Lambda
        = 0 \,, \label{eq:BSSNHamD} \\
  \mathcal{M}_i &=&
        \tilde{\gamma}^{mn} \partial_m \tilde{A}_{ni}
        - \tilde{\Upgamma}^m \tilde{A}_{mi}
        - \tilde{\gamma}^{ml} \tilde{\Upgamma}^n_{im} \tilde{A}_{nl}
        - \frac{D-2}{D-1} \partial_i K
        - \frac{D-1}{2\chi} \tilde{A}^m{}_i \partial_m \chi \nonumber \\
     && + n \tilde{\gamma}^{ww} \left( \frac{\tilde{A}_{iz}
          - \delta_{iz} \tilde{A}_{ww}}{z}
          - \tilde{\Upgamma}^m_{ww} \tilde{A}_{mi}
          - \frac{1}{2} \tilde{\gamma}^{ww} \tilde{A}_{ww}
          \partial_i \tilde{\gamma}_{ww} \right)
        -8\pi j_i = 0\,.
        \label{eq:BSSNmomD}
\end{eqnarray}
and the BSSN evolution equations
(\ref{eq:BSSNdtchi})-(\ref{eq:BSSNdtGamma}) are now written as
\begin{eqnarray}
  \partial_t \chi &=&
        \beta^m \partial_m \chi
        + \frac{2}{D-1} \chi \left(
          \alpha K - \partial_m \beta^m - n \frac{\beta^z}{z} \right)\,,
        \label{eq:dtchiD} \\
  \partial_t \tilde{\gamma}_{ij} &=&
        \beta^m \partial_m \tilde{\gamma}_{ij}
        + 2\tilde{\gamma}_{m(i} \partial_{j)} \beta^m
        - \frac{2}{D-1} \tilde{\gamma}_{ij} \left( \partial_m \beta^m
          +n \frac{\beta^z}{z} \right)
        - 2\alpha \tilde{A}_{ij}\,,
        \label{eq:dtgammaD} \\
  \partial_t \tilde{\gamma}_{ww} &=&
        \beta^m \partial_m \tilde{\gamma}_{ww}
        - \frac{2}{D-1} \tilde{\gamma}_{ww}
          \left( \partial_m \beta^m - d \frac{\beta^z}{z} \right)
        -2\alpha \tilde{A}_{ww}
        \,,
        \label{eq:dtgammawwD} \\
  \partial_t K &=&
        \beta^m \partial_m K
        - \chi \tilde{\gamma}^{mn} D_m D_n \alpha
        + \alpha \tilde{A}^{mn} \tilde{A}_{mn}
        + \frac{1}{D-1} \alpha K^2 \nonumber \\
     && + n \tilde{\gamma}^{ww} \left( \alpha
          \frac{\tilde{A}_{ww}^2}{\tilde{\gamma}_{ww}}
          - \chi D_w D_w \alpha \right)
        + \frac{2}{D-2}\alpha \left\{ 4\pi[S+(D-3)\rho] - \Lambda\right\}
        \,, \label{eq:dtKD} \\
  \partial_t \tilde{A}_{ij} &=&
        \beta^m \partial_m \tilde{A}_{ij}
        + 2\tilde{A}_{m(i} \partial_{j)} \beta^m
        - \frac{2}{D-1} \tilde{A}_{ij} \left(
          \partial_m \beta^m + n \frac{\beta^z}{z} \right)
        + \alpha K\tilde{A}_{ij} \nonumber \\
     && - 2\alpha \tilde{\gamma}^{mn}\tilde{A}_{im} \tilde{A}_{jn}
        + \chi \left[ \alpha (\mathcal{R}_{ij}-8\pi S_{ij})
          - D_i D_j \alpha \right]^{\rm TF}
        \,, \label{eq:dtAD} \\
  \partial_t \tilde{A}_{ww} &=&
        \beta^m \partial_m \tilde{A}_{ww}
        - \frac{2}{D-1} \tilde{A}_{ww}
          \left( \partial_m \beta^m - d\frac{\beta^z}{z}\right)
        + \alpha \tilde{A}_{ww}(K - 2\tilde{\gamma}^{ww} \tilde{A}_{ww})
        \nonumber \\
     && + \chi \left[ \alpha (\mathcal{R}_{ww} -8\pi S_{ww})
          - D_w D_w \alpha \right]^{\rm TF}
        \,, \label{eq:dtAwwD}
%        \\
\end{eqnarray}
\begin{eqnarray}
  \partial_t \tilde{\Upgamma}^i &=&
        \beta^m \partial_m \tilde{\Upgamma}^i
        + \frac{2}{D-1} \tilde{\Upgamma}^i \left(
          \partial_m \beta^m + n \frac{\beta^z}{z} \right)
        + \tilde{\gamma}^{mn} \partial_m \partial_n \beta^i
        + \frac{D-3}{D-1} \tilde{\gamma}^{im} \partial_m \partial_n \beta^n
        \nonumber \\
     && - \tilde{\Upgamma}^m \partial_m \beta^i
        + n \tilde{\gamma}^{ww} \left(
          \frac{\partial_z \beta^i}{z} - \delta^i{}_z \frac{\beta^z}{z^2}
          \right)
        + \frac{D-3}{D-1} n \left( \tilde{\gamma}^{im}
          \frac{\partial_m \beta^z}{z} - \tilde{\gamma}^{iz}
          \frac{\beta^z}{z^2} \right)
        \nonumber \\
     && - \tilde{A}^{im} \left[ (D-1) \alpha \frac{\partial_m \chi}{\chi}
          + 2\partial_m \alpha \right]
        + 2\alpha \left( \tilde{\Upgamma}^i_{mn} \tilde{A}^{mn}
        + n \tilde{\Upgamma}^i_{ww} \tilde{A}^{ww} \right)
        - 16\pi \frac{\alpha}{\chi} j_i
        \nonumber \\
     && - 2 \frac{D-2}{D-1} \alpha \tilde{\gamma}^{im} \partial_m K
        - \sigma \left[
        \left( \partial_m \beta^m + n \frac{\beta^z}{z} \right)
        \left( \tilde{\Upgamma}^i - \tilde{\gamma}^{mn}
          \tilde{\Upgamma}^i_{mn} - n \tilde{\gamma}^{ww}
        \tilde{\Upgamma}^i_{ww} \right) \right]\,.
        \nonumber \\
        \label{eq:dtGammaD}
\end{eqnarray}
These equations contain a number of auxiliary expressions which are given
in terms of the fundamental BSSN variables by Eq.~(\ref{eq:Gammaiww}) as well as
\begin{eqnarray}
  D_i D_j \alpha &=&
        \partial_i \partial_j \alpha
        -\tilde{\Upgamma}^m_{ji}\partial_m \alpha
        + \frac{1}{2\chi} (\partial_i \chi \partial_j \alpha
          + \partial_j \chi \partial_i \alpha)
        - \frac{\tilde{\gamma}_{ij}}{2\chi}
          \tilde{\gamma}^{mn} \partial_m \chi \partial_n \alpha\,,
        \\
  \left[ D_i D_j \alpha \right]^{\rm TF} &=&
        D_i D_j \alpha
        - \frac{1}{D-1} \tilde{\gamma}_{ij} \left(
          \tilde{\gamma}^{mn} D_m D_n \alpha
          + n \tilde{\gamma}^{ww} D_w D_w \alpha \right) \,, \\
  D_w D_w \alpha &=&
        \left( \frac{1}{2} \tilde{\gamma}^{mn} \partial_n \tilde{\gamma}_{ww}
          + \frac{\tilde{\gamma}^{zm}}{z}\tilde{\gamma}_{ww} \right)
          \partial_m \alpha
        - \frac{1}{2\chi} \tilde{\gamma}_{ww} \tilde{\gamma}^{mn}
          \partial_m \chi~\partial_n \alpha
        \,,
        \\
  \left[ D_w D_w \alpha \right]^{\rm TF} &=&
        \frac{1}{D-1} \left(
          d D_w D_w \alpha - \tilde{\gamma}_{ww} \tilde{\gamma}^{mn}
          D_m D_n \alpha \right)\,,
        \\
  \mathcal{R}_{ij} &=& \mathcal{R}^{\chi}_{ij} + \tilde{\mathcal{R}}_{ij} \,,\\
  \mathcal{R}_{ww} &=& \mathcal{R}^{\chi}_{ww} + \tilde{\mathcal{R}}_{ww} \,,\\
  \mathcal{R}^{\chi}_{ij} &=&
          \frac{1}{2\chi} \tilde{\gamma}_{ij} \left[
          \tilde{\gamma}^{mn} \tilde{D}_m \tilde{D}_n \chi
          + n \left( \frac{1}{2} \tilde{\gamma}^{ww} \tilde{\gamma}^{mn}
          \partial_n \tilde{\gamma}_{ww}
          + \frac{\tilde{\gamma}^{mz}}{z} \right) \partial_m \chi
        \right.
        \nonumber \\
     && \left.
          -\frac{D-1}{2\chi} \tilde{\gamma}^{mn} \partial_m \chi~
          \partial_n \chi
        \right]
        + \frac{D-3}{2\chi} \left( \tilde{D}_i \tilde{D}_j \chi
          - \frac{1}{2\chi} \partial_i \chi~\partial_j \chi \right)\,,
        \\
  \mathcal{R}^{\chi}_{ww} &=&
        \frac{\tilde{\gamma}_{ww}}{2\chi} \left[
          \tilde{\gamma}^{mn} \tilde{D}_m \tilde{D}_n \chi
          + (2D-d-4)\left( \frac{1}{2} \tilde{\gamma}^{ww} \tilde{\gamma}^{mn}
            \partial_n \tilde{\gamma}_{ww} + \frac{\tilde{\gamma}^{mz}}{z}
          \right) \partial_m \chi
        \right.
        \nonumber \\
     && \left.
          - \frac{D-1}{2\chi} \tilde{\gamma}^{mn}\partial_m \chi~\partial_n \chi
        \right] \,,
        \\
  \tilde{\mathcal{R}}_{ij} &=&
        n \tilde{\gamma}^{ww} \left[
        - \frac{1}{2} \frac{\partial_z \tilde{\gamma}_{ij}}{z}
        + \frac{\delta_{z(i} \tilde{\gamma}_{j)z} - \delta_{iz} \delta_{jz}
          \tilde{\gamma}_{ww}}{z^2}
        + \frac{\tilde{\gamma}^{ww} \tilde{\gamma}_{z(j} - \delta_{z(j}}{z}
        \partial_{i)} \tilde{\gamma}_{ww}
        \right.
        \nonumber \\
     && \left.
        - \frac{1}{4} \tilde{\gamma}^{ww} \partial_i \tilde{\gamma}_{ww}~
          \partial_j \tilde{\gamma}_{ww} \right]
        -\frac{1}{2}\tilde{\gamma}^{mn} \partial_m \partial_n
          \tilde{\gamma}_{ij}
        + \tilde{\gamma}_{m(i} \partial_{j)} \tilde{\Upgamma}^m
        \nonumber \\
     && + \tilde{\Upgamma}^m \tilde{\Upgamma}_{(ij)m}
        + \tilde{\gamma}^{mn} \left[
          2\tilde{\Upgamma}^k_{m(i} \tilde{\Upgamma}_{j)kn}
          + \tilde{\Upgamma}^k_{im} \tilde{\Upgamma}_{kjn} \right] \,,
        \\
  \tilde{\mathcal{R}}_{ww} &=&
        - \frac{1}{2} \tilde{\gamma}^{mn} \partial_m \partial_n
          \tilde{\gamma}_{ww}
        + \frac{1}{2} \tilde{\gamma}^{ww} \tilde{\gamma}^{mn}
          \partial_m \tilde{\gamma}_{ww}~\partial_n \tilde{\gamma}_{ww}
        - \frac{n}{2} \tilde{\gamma}^{ww}
          \frac{\partial_z \tilde{\gamma}_{ww}}{z}
        + \tilde{\gamma}_{ww} \frac{\tilde{\Upgamma}^z}{z} \nonumber \\
     && + \frac{1}{2} \tilde{\Upgamma}^m \partial_m \tilde{\gamma}_{ww}
        + \frac{\tilde{\gamma}^{zz} \tilde{\gamma}_{ww} - 1}
          {z^2}\,, \label{eq:Rwwtilde}
        \\
  \left[ \mathcal{R}_{ij} \right]^{\rm TF} &=&
        \mathcal{R}_{ij}
        - \frac{1}{D-1} \tilde{\gamma}_{ij} \tilde{\gamma}^{mn}
          \mathcal{R}_{mn}
        - \frac{n}{D-1} \tilde{\gamma}_{ij} \tilde{\gamma}^{ww}
          \mathcal{R}_{ww}\,,
%        \\
\end{eqnarray}
\begin{eqnarray}
  \left[ \mathcal{R}_{ww} \right]^{\rm TF} &=&
        \frac{1}{D-1} \left(
          d~~\mathcal{R}_{ww} - \tilde{\gamma}_{ww} \tilde{\gamma}^{mn}
          \mathcal{R}_{mn} \right)\,.
        \label{eq:YoD}
\end{eqnarray}
The BSSN equations in this form can readily be implemented in an existing
``$d$+1'' BSSN code with the addition of merely two new field variables,
$\tilde{\gamma}_{ww}$ and $\tilde{A}_{ww}$. While the BSSN equations
acquire additional terms, the computational domain remains
$d$-dimensional. Furthermore, the entire set of
Eqs.~(\ref{eq:BSSNHamD})-(\ref{eq:YoD}) contains exclusively derivatives
in the $x^i$ directions and in time, which can be evaluated without
need of ghost zones in the extra dimensions.

There only remains one further subtlety arising from the explicit division
by $z$ in several of the terms present. Some (though not all) numerical
codes require evaluation of these expressions at $z=0$ which
makes regularization of these terms mandatory. As we show explicitly in
\ref{app:regularization}, this can be achieved for all terms,
yielding expressions that are exact in the limit $z\rightarrow 0$.
The results we discuss in Sec.~\ref{sec:application} make use of these
regularized terms on the plane $z=0$ demonstrating that this
procedure provides stable and accurate evolutions.

We conclude this section with a brief remark of the matter
terms present in (\ref{eq:BSSNHamD})-(\ref{eq:YoD}) in the form
of the projections $\rho$, $j^i$, $S_{ij}$
and $S=\chi(\tilde{\gamma}^{ij}S_{ij}+n\tilde{\gamma}^{ww}S_{ww})$
of the energy-momentum
tensor. The specific form of these terms will depend on the
physical system under consideration and will need to be evaluated
separately for each case as will the precise form of the
matter evolution equations resulting from the conservation
law $\nabla_{\A}T^{\A\B}$. Many applications of higher-dimensional
numerical relativity concern BHs and the example application
discussed in Sec.~\ref{sec:application} will be an asymptotically flat
vacuum spacetime where the matter terms and the cosmological
constant are zero.

%=============================================================================
\section{The special case of $SO(2)$ symmetry}
\label{sec:SO2}
We now return to the special situation where $d=D-2$ which corresponds
to $SO(2)$ isometry, i.e.~axisymmetric spacetimes \cite{Pretorius:2004jg}. This case is special
in that $\varphi$ components of a vector field or those components
of a tensor field $T_{\A\B}$ where exactly one index is $\varphi$,
do not necessarily vanish. The reason for this exceptional property
of $SO(2)$ symmetry is that with only one Killing vector
$\boldsymbol{\partial}_{\varphi}$, the vector
$f(x^i)\,\boldsymbol{\partial}_{\varphi}$, for an arbitrary function $f$,
trivially satisfies the symmetry as it commutes with all Killing vectors.
Thus, $SO(2)$ symmetry does not, in general, cause any tensor components
to vanish, and only a negligible amount of computational cost
and memory would be saved by explicitly inserting the modified
Cartoon terms, as derived in the previous two sections, into the
BSSN equations. We are still able, however, to capitalize on the
substantial reduction in memory and floating point operations
that arises from the dimensional reduction of the
computational domain. This is most conveniently achieved
by retaining the BSSN equations in their full $D$-dimensional form
and only using
the modified Cartoon method to fill in derivatives that cannot
be calculated directly on the computational grid.

Let us illustrate this process for $D=5$ with $SO(2)$ isometry. Due to
the symmetry, we can model such a spacetime on a three-dimensional grid
on which we store all vector and tensor components, including
those of the type $V^w$ and $T_{Iw}$, which do not vanish for $SO(2)$ symmetry.
In order to evolve the system by one time step, we need to compute
derivatives with respect to {\em all} coordinates. Derivatives with
respect to $x^i$ can be calculated on the grid using standard
methods. Derivatives with respect to $w$, on the other hand,
can be calculated on our three-dimensional grid using the modified
Cartoon method. For example, for a vector, the procedure is
\begin{equation}
  {V^I}_{,J} =
  \left(
    \begin{array}{c|c}
        & 0 \\
      \text{calculate on} \;    & 0 \\
      \text{the grid} \;     & - V^w/z \\
                                     & V^z/z
    \end{array}
  \right).
  \label{eq:VIJ}
\end{equation}
\ref{app:SO2Cartoon} lists all necessary expressions for derivatives
with $SO(2)$ symmetry.  Once all derivatives have been calculated,
we have all the information required to use the standard $D=5$ BSSN
equations (\ref{eq:BSSNdtchi}-\ref{eq:BSSNdtGamma}) without the
need for any extra terms.

Note that this method for handling $SO(2)$ symmetry can straightforwardly
be combined with the method described in Secs.~\ref{sec:cartoon},
\ref{sec:BSSNdelta}. Such a procedure can handle, for example, the
symmetry of black rings and has been applied in
\cite{Figueras:2015hkb} to speed up the exploration of the
gauge parameter space in numerical evolutions of black rings
in $D=5$. Black rings have horizons of topology \cite{Emparan:2001wn}
$S^1\times S^2$ and rotate along the $S^1$. This rotational symmetry
requires handling with the special method for $d=D-2$
because $\varphi$ components do not vanish in that case. The second symmetry
corresponding to the $S^2$, however, is amenable to the treatment
presented in Secs.~\ref{sec:cartoon} and \ref{sec:BSSNdelta}.
In practice, Refs.~\cite{Figueras:2015hkb,Figueras:2016}
first applied the latter reduction
and then the special $SO(2)$ reduction sketched in
Eq.~(\ref{eq:VIJ}).

%=============================================================================
\section{Application to a black-hole collision}
\label{sec:application}
In this section we present, as a specific example for
the efficacy of the formalism,
results from the numerical simulation of a head-on collision
of two non-spinning BHs in $D=7$ dimensions starting from rest.
A non-rotating BH in $D$ spacetime dimensions is described by
the Tangherlini\cite{Tangherlini:1963bw} solution
\begin{equation}
  ds^2 = -\left(1-\frac{\mu}{R^{D-3}} \right)dt^2
        +\left(1-\frac{\mu}{R^{D-3}}\right)^{-1} dR^2
        + R^2 d\Omega^2_{D-2}\,,
\end{equation}
where $d\Omega_{D-2}$ denotes the area element of the $(D-2)$ sphere and
the parameter $\mu$ is related to the BH mass $M$ and the
horizon radius $R_{\rm h}$ by
\begin{equation}
  \mu = \frac{16\pi M}{(D-2)\Omega_{D-2}}\,,~~~~~
  \mu = R_{\rm h}^{D-3}\,.
\end{equation}
Here, $\Omega_{D-2}$ is the surface area of the unit $(D-2)$ sphere. The Tangherlini
solution can be written in isotropic coordinates in the form
\begin{equation}
  ds^2 = -\left( \frac{4r^{D-3}-\mu}{4r^{D-3}+\mu}\right)dt^2
        +\left(1+\frac{\mu}{4r^{D-3}}\right)^{4/(D-3)}
        \left[ \sum_i (dx^i)^2 + \sum_a (dw^a)^2 \right]\,,
\end{equation}
which facilitates construction of analytic data for a snapshot of a
spacetime containing multiple BHs at the moment of time symmetry
according to the procedure of Brill and Lindquist \cite{Brill:1963yv}.
These higher-dimensional Brill-Lindquist data are given in terms of the
ADM variables by
\begin{equation}
  K_{\I \J} = 0\,,~~~
  \gamma_{\I \J} = \psi^{4/(D-3)} \delta_{\I \J}\,,~~~
  \psi = 1 + \sum_{\mathcal{A}} \frac{\mu_{\mathcal{A}}}
         {4 \left[\sum_{\K=1}^{D-1}
         (X^{\K} - X_{\mathcal{A}}^{\K})^2\right]^{(D-3)/2}} \,,\label{BLid}
\end{equation}
where the summation over $\mathcal{A}$ and $K$ extend over the multiple
BHs and spatial coordinates, respectively, and $X^{\K}_{\mathcal{A}}$
denotes the position of the $\mathcal{A}^{\rm th}$ BH.

We have implemented these initial data in the {\sc Lean} code
\cite{Sperhake:2006cy,Sperhake:2007gu}, which is based on {\sc Cactus}
\cite{Cactusweb,Allen:1999} and uses {\sc Carpet}
\cite{Carpetweb,Schnetter:2003rb} for mesh refinement.
The specific case presented here has been obtained
using $SO(4)$ symmetry, i.e.~$D=7$, $d=3$,
for a collision along the $x$ axis of two equal-mass
BHs initially separated by $7.58~R_h$, where $R_h$ is the horizon
radius associated with a single BH with $\mu=\mu_1=\mu_2$.
The computational domain consists of a set of seven refinement levels,
the innermost two centered on the BHs and the five outer ones on the origin.
We employ standard {\em moving puncture} gauge conditions
\cite{vanMeter:2006vi} [note that we use here $\beta^a=0$ in accordance with
Eq.~(\ref{eq:CartVectorZero})]
\begin{eqnarray}
  \partial_t \alpha &=& \beta^m \partial_m \alpha
        - 3\alpha K\,, \\
  \partial_t \beta^i &=& \beta^m \partial_m \beta^i
        + \frac{3}{4} \tilde{\Gamma}^i - \frac{1}{2^{1/4}R_h} \beta^i\,,
\end{eqnarray}
having initialized lapse and shift to their Minkowski values
$\alpha=1$, $\beta^i=0$.
Two simulations have been performed in octant symmetry with a grid spacing
$\Delta x = R_h / 52$ and $\Delta x = R_h / 104$, respectively, on the
innermost level, that increases by a factor of two on each consecutive level
further out.

\begin{figure}[b]
  \centering
  \includegraphics[height=200pt,clip=true]{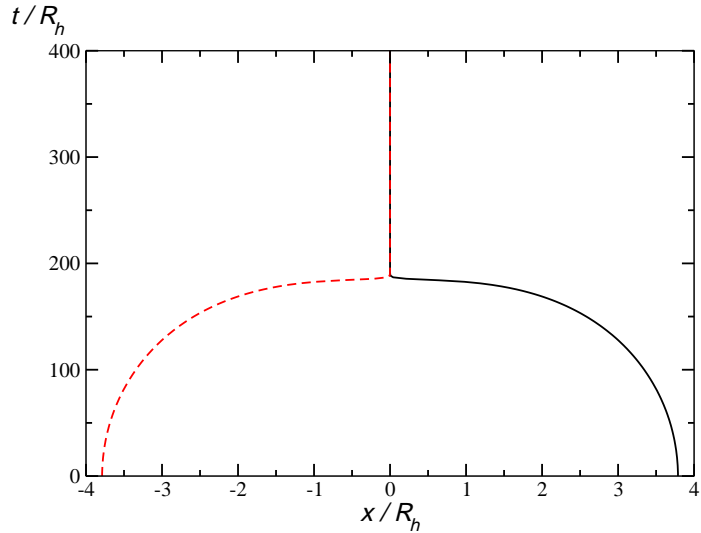}
  \caption{BH trajectories for an equal-mass head-on collision of
           two non-spinning holes initially at rest in $D=7$ dimensions.
           The collision takes place along the $x$ axis.}
  \label{fig:trajd7}
\end{figure}

\begin{figure}[b]
  \centering
  \includegraphics[height=200pt,clip=true]{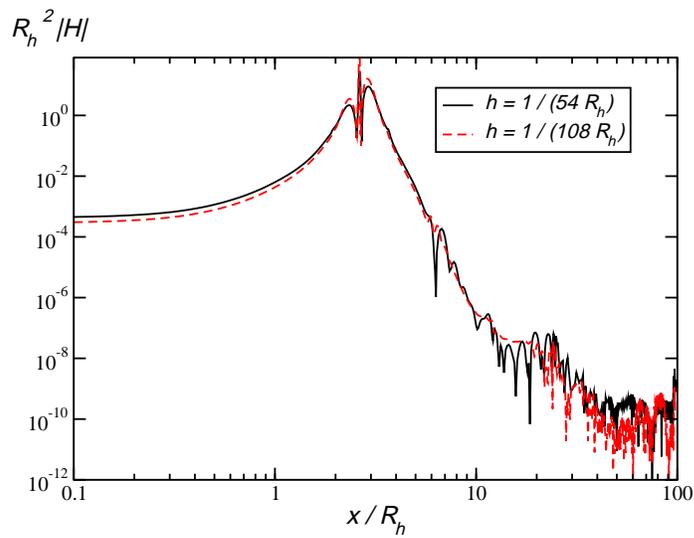}
  \caption{The Hamiltonian constraint along the collision axis obtained
           for a BH head-on collision starting from rest using resolution
           parameters $\Delta x = R_h/52$ (solid, black curve) and
           $\Delta x = R_h/104$ (dashed, red curve). The latter has been
           amplified by a factor of four corresponding to second-order
           convergence.
          }
  \label{fig:ham}
\end{figure}

Figure \ref{fig:trajd7} shows the trajectories of the two BHs
evolving in time from the initial separation through merger into a single
hole centered on the origin, obtained from the high resolution simulation
with $\Delta x = R_h/104$.
In order to check the consistency of our numerical formalism, we have also
analyzed the constraint equations for this configuration. A snapshot of
the Hamiltonian constraint (\ref{eq:BSSNHam}) along the collision
axis at evolution time $t=80~R_h$ is shown in Fig.~\ref{fig:ham}.
In this figure, the result obtained for the high resolution run has been
amplified by a factor of four expected for second-order convergence.
The overlap of the two curves demonstrates convergence at second order,
compatible with the numerical scheme that employs second and fourth-order
accurate discretization and interpolation techniques.
We have performed the same analysis for the Hamiltonian and momentum
constraints at several points in time and observe the same second-order
convergence of both constraints throughout infall and merger. Note
that only one BH is present on the computational domain (at about
$x=2.5$ in the figure) because of the octant symmetry. The other
BH is represented in this simulation through the symmetric boundary
conditions imposed at $x=0$.

%=============================================================================
\section{Conclusions}
\label{sec:conclusions}
In the presence of rotational symmetry, the Einstein equations
simplify considerably and the generation of numerical solutions
to these equations can be implemented with significant improvements
in computational cost and the required amount of computer memory.
The Cartoon method proposed in \cite{Alcubierre:1999ab}
was the first technique designed with the particular goal of
efficiently modeling axisymmetric spacetimes in 3+1 numerical
relativity. A modification, often dubbed the modified
Cartoon method \cite{Pretorius:2004jg}
used relations between tensor components
in place of spatial interpolation operations, which not only
eliminates the need of introducing a few extra grid points
in the symmetry directions, but also allows for a particularly
convenient generalization to an arbitrary number of spacetime dimensions
and number of rotational symmetries
\cite{Shibata:2010wz,Yoshino:2011zz,Yoshino:2011zza}.

In this work, we have presented
in detail the complete set of equations as obtained for the BSSN
formulation of the Einstein equations in $D$ spacetime dimensions
with $SO(D-d)$ isometry where $d \in \{1,\,2,\,\ldots,\,D-2\}$.
Furthermore, we explicitly demonstrate the presence of extra terms
for the case $d=D-2$, where the symmetry condition allows for
a wider class of components of tensor densities to remain non-zero.
Finally, we have compiled a list of terms involving division by
the quasi-radial coordinate (the $z$ direction in our case) and
illustrate how all irregularities at the origin $z=0$ can be cured
through equivalence with manifestly regular expressions. Even though we
used the BSSN formulation for our discussion, the recipes detailed here
can be applied straightforwardly to other popular formulations
of the Einstein equations such as the generalized harmonic gauge
\cite{Garfinkle:2001ni,Pretorius:2004jg} or the conformal $Z4$
\cite{Alic:2011gg,Weyhausen:2011cg} systems.

As an example, we have presented results from a head-on collision
from rest of two equal-mass, non-spinning BHs in $D=7$ spacetime
dimensions. Following a rather slow acceleration phase, due to the
rapid diminishing of the gravitational force with distance, the
two BHs merge and we observe second-order convergence of the
constraints. This confirms in yet another type of application
the remarkable robustness observed for the modified Cartoon method
in applications to spinning BHs\cite{Shibata:2010wz} or high-energy
collisions in $D=5$ \cite{Okawa:2011fv}. This seemingly superior
robustness as compared with the method of reduction by isometry developed in
\cite{Zilhao:2010sr} is, at present, empirical but merits
further investigation at the analytic level.

%=============================================================================
\section*{Acknowledgments}
U.S. is supported by the H2020 ERC Consolidator Grant ``Matter and
strong-field gravity: New frontiers in Einstein's theory'' grant
agreement No. MaGRaTh--646597, the H2020-MSCA-RISE-2015 Grant
No. StronGrHEP-690904, the STFC Consolidator Grant No. ST/L000636/1,
the SDSC Comet and TACC Stampede clusters through NSF-XSEDE Award
Nos.~PHY-090003, the Cambridge High Performance
Computing Service Supercomputer Darwin using Strategic Research
Infrastructure Funding from the HEFCE and the STFC, and DiRAC's Cosmos
Shared Memory system through BIS Grant No.~ST/J005673/1 and STFC Grant
Nos.~ST/H008586/1, ST/K00333X/1. P.F. and S.T. are supported by the H2020 ERC Starting Grant ``New frontiers in numerical general relativity'' grant agreement No. NewNGR-639022. P.F. is also supported by a Royal Society University Research Fellowship. W.G.C. and M.K. are supported by STFC studentships.

%=============================================================================
\appendix
\section{Cartesian components in $SO(D-d)$ symmetry} \label{app:cartoon}
We present here the list of all modified cartoon expressions for
the case of $SO(D-d)$ symmetry with $d<D-2$. The index range for
early Latin indices is $a,\,b,\,\ldots = d+1,\,\ldots,\,D-1$ and
for middle Latin indices $i,\,j,\,\ldots=1,\, \ldots \, d$.
Furthermore, an index $z$ denotes the
coordinate $z$ while the index $w$ only appears in the
tensor component $T_{ww}$ which represents the additional
function that needs to be evolved numerically in addition to
the $T_{ij}$. For example, the spacetime metric is fully described
by the components $g_{\alpha \beta}$, $\alpha,\,\beta = 0,\,1,\, \ldots \, d$,
plus one additional field $g_{ww}$. For arbitrary scalar, vector and
tensor densities $\Psi$, $V^{\A}$ and $T_{\A\B}$, the expressions are
\begin{eqnarray}
  0 &=& \partial_a \Psi = \partial_i \partial_a \Psi\,,
        \label{eq:CartScalar1} \\
  \partial_a \partial_b \Psi &=& \delta_{ab} \frac{\partial_z \Psi}{z}\,,
        \label{eq:CartScalar2}\\
  0 &=& V^a = \partial_i V^a = \partial_a V^i = \partial_a \partial_b V^c\,,
        \label{eq:CartVectorZero}\\
  \partial_a V^b &=& \delta_a{}^b \frac{V^z}{z}\,, \\
  \partial_i \partial_a V^b &=& \delta^b{}_a \left( \frac{\partial_i V^z}{z}
        - \delta_{iz} \frac{V^z}{z^2}\right)\,,
        \label{eq:CartVector1} \\
  \partial_a \partial_b V^i &=& \delta_{ab} \left(
        \frac{\partial_z V^i}{z}-\delta^i{}_z \frac{V^z}{z^2} \right)\,,
        \label{eq:CartVector2}
        \\[10pt]
  0 &=& T_{ia} = \partial_a T_{bc} = \partial_i \partial_a T_{bc}
        = \partial_a \partial_b T_{ic} = \partial_a T_{ij}
        = \partial_i \partial_a T_{jk}\,,\label{eq:Cart2TensorZero}\\
  T_{ab} &=& \delta_{ab} T_{ww}\,, \\
  \partial_a \partial_b T_{cd} &=&
        (\delta_{ac}\delta_{bd} + \delta_{ad}\delta_{bc})
        \frac{T_{zz}-T_{ww}}{z^2} + \delta_{ab}\delta_{cd}
        \frac{\partial_z T_{ww}}{z}\,,
        \label{eq:CartTensor2} \\
  \partial_a T_{ib} &=& \delta_{ab} \frac{T_{iz}-\delta_{iz}T_{ww}}{z}\,,
        \label{eq:CartTensor3} \\
  \partial_i \partial_a T_{jb} &=& \delta_{ab} \left(
        \frac{\partial_i T_{jz}-\delta_{jz} \partial_i T_{ww}}{z}
        -\delta_{iz} \frac{T_{jz}-\delta_{jz}T_{ww}}{z^2} \right)\,,
        \label{eq:CartTensor4} \\
  \partial_a \partial_b T_{ij} &=& \delta_{ab} \left(
        \frac{\partial_z T_{ij}}{z}
        - \frac{\delta_{iz} T_{jz}+\delta_{jz}T_{iz}-2\delta_{iz}\delta_{jz}
                T_{ww}}{z^2} \right)\,.
        \label{eq:CartTensor5}
\end{eqnarray}
%

%=============================================================================

\section{Regularization at $z=0$ for $d<D-2$}
\label{app:regularization}
The presence of $z$ in the denominator of several terms in the system
of Eqs.~(\ref{eq:BSSNHamD})-(\ref{eq:YoD}) merely
arises from the quasi-radial nature of the coordinate $z$ and
can be handled straightforwardly in analogy to the treatment of
the origin in spherical or axisymmetry.

We will present the regularized terms needed in the generic $SO(D-d)$ symmetry; however, it should be noted that terms involving the inverse metric become much more complicated for a large $d$, and so we will also explicitly show these terms for the most common case, $d=3$.

We first require that all components expressed in a fully
Cartesian set of coordinates are regular. A well known consequence of
this assumption is that tensor density components containing
an odd (even) number of radial, i.e.~$z$, indices contain
only odd (even) powers of $z$ in a series expansion around $z=0$.
The same holds for quantities derived from tensors and densities
such as the BSSN variable $\tilde{\Upgamma}^{i}$.

Next, we consider the inverse metric which we obtain through
inversion of the matrix equation (\ref{eq:gammamatrix}).
By constructing the cofactor matrix and dividing by the
determinant,
we obtain, for $d=3$
\begin{equation}
  \begin{array}{lll}
    \tilde{\gamma}^{xx} = \tilde{\gamma}_{ww}^n
        \frac{\tilde{\gamma}_{yy} \tilde{\gamma}_{zz} - \tilde{\gamma}_{yz}^2}
             {\det\tilde{\gamma}_{\I\J}}
    \,,~&
    \tilde{\gamma}^{xy} = \tilde{\gamma}_{ww}^n
        \frac{\tilde{\gamma}_{yz} \tilde{\gamma}_{xz}
              - \tilde{\gamma}_{xy} \tilde{\gamma}_{zz}}
             {\det\tilde{\gamma}_{\I\J}}
    \,,~&
    \tilde{\gamma}^{xz} = \tilde{\gamma}_{ww}^n
        \frac{\tilde{\gamma}_{xy} \tilde{\gamma}_{yz}
              - \tilde{\gamma}_{xz} \tilde{\gamma}_{yy}}
             {\det\tilde{\gamma}_{\I\J}}
    \,, \\[10pt]
    \cdots &
    \tilde{\gamma}^{yy} = \tilde{\gamma}_{ww}^n
        \frac{\tilde{\gamma}_{xx} \tilde{\gamma}_{zz} - \tilde{\gamma}_{xz}^2}
             {\det\tilde{\gamma}_{\I\J}}
    \,, &
    \tilde{\gamma}^{yz} = \tilde{\gamma}_{ww}^n
        \frac{\tilde{\gamma}_{xy} \tilde{\gamma}_{xz}
              - \tilde{\gamma}_{xx} \tilde{\gamma}_{yz}}
             {\det\tilde{\gamma}_{\I\J}}
    \,, \\[10pt]
    \cdots & \cdots &
    \tilde{\gamma}^{zz} = \tilde{\gamma}_{ww}^n
        \frac{\tilde{\gamma}_{xx} \tilde{\gamma}_{yy} - \tilde{\gamma}_{xy}^2}
             {\det\tilde{\gamma}_{\I\J}}
        \,.
  \end{array}
  \label{eq:BSSNinvgamma}
\end{equation}
Next, we recall that the BSSN metric has unit determinant, so that
\begin{eqnarray}
  1 = \det \tilde{\gamma}_{\I\J} &=&
        \tilde{\gamma}_{ww}^n
        ( \tilde{\gamma}_{xx}\tilde{\gamma}_{yy}\tilde{\gamma}_{zz}
         +2\tilde{\gamma}_{xy}\tilde{\gamma}_{xz}\tilde{\gamma}_{yz}
         -\tilde{\gamma}_{xx}\tilde{\gamma}_{yz}^2
         -\tilde{\gamma}_{yy}\tilde{\gamma}_{xz}^2
         -\tilde{\gamma}_{zz}\tilde{\gamma}_{xy}^2 ) \nonumber \\
    &\overset{*}{=}& \tilde{\gamma}_{ww}^n \tilde{\gamma}_{zz}
        ( \tilde{\gamma}_{xx}\tilde{\gamma}_{yy}
         -\tilde{\gamma}_{xy}^2 )\,,
\end{eqnarray}
where we introduced the symbol ``$\,\overset{*}{=}\,$'' to denote
equality in the limit $z\rightarrow 0$. The components
for the inverse BSSN metric in Eq.~(\ref{eq:BSSNinvgamma}) simplify accordingly.

For a general $d$ we know that the matrix takes the form given in Eq (\ref{eq:gammamatrix}). 
Then, denoting the cofactor matrix for a given element of $\tilde\gamma_{\I\J}$ as $C_{\I\J}$, the inverse BSSN metric components are
(note that the metric is symmetric, so that $C_{\I \J} = C_{\J \I}$)
\begin{equation}
  \begin{array}{cccc}
  \tilde{\gamma}^{x^1x^1} =
  \frac{C_{x^1x^1}}{\det{\tilde{\gamma}_{\I\J}}}
  \,,~&
  \cdots
  \,,~&
  \tilde{\gamma}^{x^1x^{d-1}} =
  \frac{C_{x^1x^{d-1}}}{\det{\tilde{\gamma}_{\I\J}}}
  \,,~&
  \tilde{\gamma}^{x^1z} =
  \frac{C_{x^1z}}{\det{\tilde{\gamma}_{\I\J}}}
  \,, \\[10pt]
  \vdots &
  \ddots &
  \vdots &
  \vdots \\[10pt]
  \vdots &
  \cdots & 
  \tilde{\gamma}^{x^{d-1}x^{d-1}} =
  \frac{C_{x^{d-1}x^{d-1}}}{\det{\tilde{\gamma}_{\I\J}}}, & 
  \tilde{\gamma}^{x^{d-1}z} =
        \frac{C_{x^{d-1}z}}{\det{\tilde{\gamma}_{\I\J}}} \\[10pt]
  \cdots & 
  \cdots &
  \cdots & 
  \tilde{\gamma}^{zz} =
  \frac{C_{zz}}{\det{\tilde{\gamma}_{\I\J}}}
  \,.
  \end{array}
  \label{eq:BSSNinvgammadelta}
\end{equation}
Again, in the BSSN case
$\det\tilde{\gamma}_{\I\J} =1$, and the inverse metric element is simply the cofactor of that element. For simplicity, we will use indices $\hat{i}$ in place of
$x^{\hat{i}}$ in the remainder of this section, so that, for example
$C_{12} \equiv C_{x^1x^2}$, $C_{1z}\equiv C_{x^1 z}$ etc. When used without
a caret, the lower case Latin indices $i,\,j,\,\ldots$ also include
the $z$ component.

If we denote the upper-left quadrant of the matrix in Eq
(\ref{eq:gammamatrix}) as the matrix $M_{ij}$, then we can write
the cofactor of an element in this upper-left quadrant as
\begin{equation}
  C_{ij}=(-1)^{i+j}\tilde{\gamma}_{ww}^n \det(M_{kl\{k\ne j, l\ne i\}})\,.
  \label{eq:cofactor}
\end{equation}
Here, the notation $\det(M_{kl\{k\ne j, l\ne i\}})$ denotes the
determinant of the matrix obtained by crossing out the $j^{\rm th}$
row and $i^{\rm th}$ column. Likewise, we may add further inequalities
inside the braces to denote matrices obtained by crossing out more
than one row and column.

The next regularity condition we require our spacetime to satisfy is the
absence of a conical singularity at $z=0$.
In polar coordinates $(\rho,\varphi)$ constructed as in
Sec.~\ref{sec:components}, this condition can be expressed as
$\tilde{\gamma}_{\varphi \varphi}=\rho^2 \tilde{\gamma}_{\rho \rho}$
which translates into the conditions
\begin{equation}
  \tilde{\gamma}_{zz}-\tilde{\gamma}_{ww}
        \overset{*}{=}\mathcal{O}(z^2)\,,~~~~~~~
  \tilde{\gamma}^{zz}-\tilde{\gamma}^{ww}
        \overset{*}{=}\mathcal{O}(z^2)\,,~~~
\end{equation}
in Cartesian coordinates. By taking the time derivative of these
relations and combining these with Eqs.~(\ref{eq:dtgammaD}),
(\ref{eq:dtgammawwD}), we
obtain an analogous relation for the traceless extrinsic
curvature,
\begin{equation}
  \tilde{A}_{zz}-\tilde{A}_{ww} \overset{*}{=} \mathcal{O}(z^2)\,.
\end{equation}
We thus arrive at the following list of regularized terms
valid in the limit $z\rightarrow 0$.
\begin{list}{\rm{{\bf(\arabic{count})}}}{\usecounter{count}
             \labelwidth0.5cm \leftmargin0.7cm \labelsep0.2cm \rightmargin0cm
             \parsep0.5ex plus0.2ex minus0.1ex \itemsep0ex plus0.2ex}
\item
  By expanding $\beta^z = b_1 z + b_3 z^3 + \ldots$, and
  likewise for $\tilde{\Gamma}^z$ and
  $\partial_z \tilde{\gamma}_{ww}$, we obtain
  \begin{equation} \label{eq:regPoint1}
    \frac{\beta^z}{z} \overset{*}{=} \partial_z \beta_z\,,~~~~~~
    \frac{\tilde{\Gamma}^z}{z} \overset{*}{=}
        \partial_z \tilde{\Gamma}^z\,,~~~~~~
    \frac{\partial_z \tilde{\gamma}_{ww}}{z}
        \overset{*}{=} \partial_z \partial_z \tilde{\gamma}_{ww}\,,
  \end{equation}
  and likewise for $\alpha$ or $\chi$ in place of
  $\tilde{\gamma}_{ww}$ in the last expression.
\item
  We express the inverse metric components through their cofactors,
  given for arbitrary $d$ by Eq.~(\ref{eq:cofactor}), and then
  apply the same trading of divisions by $z$ for derivatives
  as done for $\beta^z/z$, to obtain
  \begin{equation}
        \frac{\delta^i{}_z - \tilde{\gamma}^{zi} \tilde{\gamma}_{ww}}{z}
        \overset{*}{=} \left\{
        \begin{array}{ll} \displaystyle
    \sum_{\hat{m}=1}^{d-1}
    (-1)^{\hat{m}+\hat{i}}\partial_z(\tilde{\gamma}_{{\hat{m}} z})
    \tilde{\gamma}_{ww}^{n+1} \det(M_{jl \{j \ne z, j\ne \hat{m},
    l \ne i, l \ne z \}})~~&
        \text{if }i = \hat{i} \\[20pt]
        0 & \text{if }i = z
        \end{array}
        \right.\,.
        \label{eq:reg2}
  \end{equation}
  Here, as well in items {\bf (5)} and {\bf (9)} below,
  we formally set $\det(M_{jl\{j\ne z,j\ne \hat{m},l\ne i,l\ne z\}})=1$
  for the case $d=2$ where no entries would be left in the
  matrix after crossing out two rows and columns. For $d=1$,
  the case $i=\hat{i}$ does not arise which obviates the need to
  evaluate the determinant.
  For the case $d=3$, the expression (\ref{eq:reg2}) becomes
  \begin{equation}
    \frac{\delta^i{}_z - \tilde{\gamma}^{zi} \tilde{\gamma}_{ww}}{z}
          \overset{*}{=} \left\{
      \begin{array}{ll}
          \tilde{\gamma}_{ww}^{n+1}
            (\tilde{\gamma}_{yy} \partial_z \tilde{\gamma}_{xz}
             - \tilde{\gamma}_{xy} \partial_z \tilde{\gamma}_{yz})~~~~ &
            \text{if }i = x \\[10pt]
          \tilde{\gamma}_{ww}^{n+1}
            (\tilde{\gamma}_{xx} \partial_z \tilde{\gamma}_{yz}
             - \tilde{\gamma}_{xy} \partial_z \tilde{\gamma}_{xz})~~~~ &
            \text{if }i = y \\[10pt]
          0 & \text{if }i = z
      \end{array}
      \right.\,.
  \end{equation}
\item
  Expanding $\beta^{\hat{i}}=b_0 + b_2 z^2+\ldots$ and
  $\beta^z=b_1 z + b_3 z^3+\ldots$\,, we trade two divisions by $z$
  for a second derivative and obtain
  \begin{equation}
    \frac{\partial_z \beta^i}{z} - \delta^i{}_z \frac{\beta^z}{z^2}
          = \left\{
      \begin{array}{ll}
          \partial_z \partial_z \beta^{\hat{i}}~~~&
          \text{if } i = \hat{i} \\[10pt]
          0 & \text{if }i = z
      \end{array}
      \right.\,.
  \end{equation}
\item
  We rewrite the term
  \begin{equation}
    \frac{\tilde{\gamma}^{im} \partial_m \beta^z}{z}
          - \tilde{\gamma}^{iz} \frac{\beta^z}{z^2}
          = \tilde{\gamma}^{im} \left(
            \frac{\partial_m \beta^z}{z}
            - \delta^z{}_m \frac{\beta^z}{z^2} \right)\,,
  \end{equation}
  and expand $\beta^z=b_1z + b_3z^3+\ldots$ which leads to
  \begin{equation}
    \frac{\partial_m \beta^z}{z}
            - \delta^z{}_m \frac{\beta^z}{z^2} =
          \left\{
      \begin{array}{ll}
          \partial_{\hat{m}} \partial_z \beta^z~~~&
          \text{if }m = \hat{m} \\[10pt]
          0 & \text{if }m = z   \label{eq:regPoint4}
      \end{array}
      \right. \,.
  \end{equation}
\item
  Similarly to Eq.~(\ref{eq:reg2}), we find for general $d$ that
  \begin{eqnarray}
    \frac{\tilde{\gamma}^{zm}}{z} \partial_m \alpha &=&
    \sum_{\hat{m}=1}^{d-1}
    \sum_{\hat{i}=1}^{d-1}
    (-1)^{\hat{m}+\hat{i}-1}\partial_z(\tilde{\gamma}_{{\hat{m}} z})
    \tilde{\gamma}_{ww}^{n} \det(M_{jl \{j \ne z, j\ne \hat{m}, l \ne
    \hat{i}, l \ne z \}}) \partial_{\hat{i}}\alpha \nonumber\\
    && + \tilde{\gamma}^{zz} \partial_z \partial_z \alpha\,,
  \end{eqnarray}
  where again we formally set
  $\det(M_{jl\{j\ne z,j\ne \hat{m},l\ne i,l\ne z\}})=1$
  for the case $d=2$; cf.~item {\bf (2)} above.
  For $d=3$, we obtain
  \begin{eqnarray}
    \frac{\tilde{\gamma}^{zm}}{z} \partial_m \alpha &=&
          \tilde{\gamma}_{ww}^n \left[
          \left(
            \tilde{\gamma}_{xy} \partial_z \tilde{\gamma}_{yz}
            - \tilde{\gamma}_{yy} \partial_z \tilde{\gamma}_{xz}
          \right)
          \partial_x \alpha
          + \left(
          \tilde{\gamma}_{xy} \partial_z \tilde{\gamma}_{xz}
            - \tilde{\gamma}_{xx} \partial_z \tilde{\gamma}_{yz}
          \right)
          \partial_y \alpha \right] \nonumber
    \\
       && + \tilde{\gamma}^{zz} \partial_z \partial_z \alpha\,,
  \end{eqnarray}
  and likewise for $\chi$ in place of $\alpha$.
\item
  Using $\tilde{A}_{zz}-\tilde{A}_{ww}= \mathcal{O}(z^2)$, we obtain
  \begin{equation}
    \frac{\tilde{A}_{iz} - \delta_{iz} \tilde{A}_{ww}}{z}
          = \left\{
      \begin{array}{ll}
          \partial_z \tilde{A}_{\hat{i}z}~~~~~&
          \text{if }i = \hat{i}\\[10pt]
          0 & \text{if }i = z
      \end{array}
      \right.
      \,.
      \label{eq:reg07}
  \end{equation}
\item
  Using $\tilde{\gamma}_{zz}-\tilde{\gamma}_{ww} = \mathcal{O}(z^2)$
  and trading a division by $z$ for a $z$ derivative, we find
  \begin{equation}
    -\frac{1}{2} \frac{\partial_z \tilde{\gamma}_{ij}}{z}
          + \frac{\delta_{z(i} \tilde{\gamma}_{j)z} - \delta_{iz}
            \delta_{jz} \tilde{\gamma}_{ww}}{z^2}
          = \left\{
      \begin{array}{ll}
          -\frac{1}{2} \partial_z \partial_z
            \tilde{\gamma}_{\hat{i}\hat{j}}~~~~~&
            \text{if }(i,j)=(\hat{i},\hat{j}) \\[10pt]
          0 & \text{if }(i,j)=(\hat{i},z) \text{ or }(z,\hat{j}) \\[10pt]
          -\frac{1}{2} \partial_z \partial_z \tilde{\gamma}_{ww} &
            \text{if }(i,j)=(z,z)
      \end{array}
      \right.\,.
  \end{equation}
\item
  Using $\tilde{\gamma}^{ww} \tilde{\gamma}_{zz}-1
  =\tilde{\gamma}^{ww}(\tilde{\gamma}_{zz}-\gamma_{ww})
  = \tilde{\gamma}^{ww} \mathcal{O}(z^2)$ and
  $\tilde{\gamma}_{z\hat{i}}/z = \partial_z \tilde{\gamma}_{z\hat{i}}$,
  we can rewrite
  \begin{equation}
    \frac{\tilde{\gamma}^{ww} \tilde{\gamma}_{z(j} - \delta_{z(j}}{z}
          \partial_{i)} \tilde{\gamma}_{ww} = \left\{
      \begin{array}{ll}
          \tilde{\gamma}^{ww} \partial_z \tilde{\gamma}_{z(\hat{j}}
          \partial_{\hat{i})} \tilde{\gamma}_{ww}~~~~~&
          \text{if }(i,j)=(\hat{i},\hat{j}) \\[10pt]
          0 & \text{if }(i,j)=(\hat{i},z)\text{ or }(z,\hat{j}) \\[10pt]
          0 & \text{if }(i,j)=(z,z)
      \end{array}
    \right. \,.
  \end{equation}
\item
  The term $(\tilde{\gamma}^{zz} \tilde{\gamma}_{ww}-1) /z^2$
  requires slightly more work and we describe its derivation here
  in a little more detail. We first rewrite this term in the form
  \begin{equation}
    \frac{\tilde{\gamma}^{zz} \tilde{\gamma}_{ww}-1}{z^2}
        = -\tilde{\gamma}^{zz} \frac{\frac{1}{\tilde{\gamma}^{zz}}
          -\tilde{\gamma}_{ww}}{z^2}\,,
    \label{eq:reg9}
  \end{equation}
  and express the inverse metric component $\tilde{\gamma}^{zz}$
  in terms of the corresponding cofactor matrix component and
  the determinant as
  \begin{equation}
    \frac{1}{{\tilde{\gamma}^{zz}}}=\frac{\det
        {\tilde{\gamma}_{\I\J}}}{C_{zz}}
        =\frac{\tilde{\gamma}_{zz}C_{zz}}{C_{zz}}+
        \frac{\sum_{\hat{i}=1}^{d-1}
        \tilde{\gamma}_{z{\hat{i}}}C_{zx_{\hat{i}}}}{C_{zz}}.
  \end{equation}
  Note that these expressions are all valid for arbitrary values of $z$
  and we are not yet using the BSSN condition $\det{\tilde\gamma}_{\I\J}=1$.
  We can now plug this relation into Eq.~(\ref{eq:reg9}). We then
  trade divisions by $z$ for derivatives with respect to $z$, bearing
  in mind that $\tilde{\gamma}_{zz}=\tilde{\gamma}_{ww}+\mathcal{O}(z^2)$
  and find
  \begin{eqnarray}
    &&\frac{\tilde{\gamma}^{zz}\tilde{\gamma}_{ww}-1}{z^2}\overset{*}{=}
        \frac{\tilde{\gamma}^{zz}}{2}
        \left(\partial_z\partial_z\tilde{\gamma}_{ww}
              -\partial_z\partial_z\tilde{\gamma}_{zz}\right)
        \\
    && ~~~~~~ +\tilde{\gamma}^{zz}\sum_{\hat{i}=1}^{d-1}
        \sum_{\hat{j}=1}^{d-1}
        (-1)^{\hat{i}+\hat{j}}\,
        \tilde{\gamma}_{ww}^n
        \frac{\partial_z \tilde{\gamma}_{z\hat{i}}\,
              \partial_z \tilde{\gamma}_{\hat{j}z}}{C_{zz}}
        \det(M_{kl\{k\ne z,k\ne\hat{j},l\ne \hat{i},l\ne z\}})\,. \nonumber
  \end{eqnarray}
  Again, we formally set $\det(M_{jl\{j\ne z,j\ne \hat{m},l\ne i,l\ne z\}})=1$
  for the case $d=2$; cf.~item {\bf (2)} above.
  Finally we use $1=\det\tilde{\gamma}_{\I\J}\Rightarrow
  C_{zz}=\tilde{\gamma}^{zz}$
  to obtain
  \begin{eqnarray}
  \frac{\tilde\gamma^{zz}\tilde\gamma_{ww}-1}{z^2}&\overset{*}{=}&
        \frac{\tilde{\gamma}^{zz}}{2}
        \left(\partial_z\partial_z\tilde\gamma_{ww}
              -\partial_z\partial_z\tilde\gamma_{zz}\right)
        \\
     && +\sum_{\hat{i}=1}^{d-1}
        \sum_{\hat{j}=1}^{d-1}(-1)^{\hat{i}+\hat{j}}\,
        \tilde{\gamma}_{ww}^n
        (\partial_z\tilde{\gamma}_{z{\hat{i}}})\,
        \partial_z\tilde{\gamma}_{z\hat{j}}
        \det(M_{kl\,\{k\ne z,k\ne\hat{j},l\ne\hat{i},l\ne z\}})\,.
        \nonumber
  \end{eqnarray}
  For the case $d=3$ this reduces to:
  \begin{eqnarray}
    \frac{\tilde{\gamma}^{zz}\tilde{\gamma}_{ww}-1}{z^2}
        &\overset{*}{=}& \frac{\tilde{\gamma}^{zz}}{2}
        \partial_z \partial_z(\tilde{\gamma}_{ww}-\tilde{\gamma}_{zz})
        - \tilde{\gamma}_{ww}^n \left[
        2\tilde{\gamma}_{xy}(\partial_z \tilde{\gamma}_{xz})
        \partial_z \tilde{\gamma}_{yz}
        -\tilde{\gamma}_{xx}(\partial_z\tilde{\gamma}_{yz})^2
        \right.
        \nonumber \\
     && \left. -\tilde{\gamma}_{yy} (\partial_z \tilde{\gamma}_{xz})^2
        \right]\,.
  \end{eqnarray}
\end{list}
%

%=============================================================================

\section{Cartesian components in $SO(2)$ symmetry} \label{app:SO2Cartoon}
The general case of $SO(2)$ symmetry requires some modifications
to the expressions given in \ref{app:cartoon}. Here we list these
necessary changes.
Recall that
lower case Latin indices with a caret range from $1,...,D-3$,
since $d=D-2$.

The expressions for scalars \eqref{eq:CartScalar1} and \eqref{eq:CartScalar2} remain unchanged.
For vectors, Eq.~\eqref{eq:CartVectorZero}
no longer holds in $SO(2)$ symmetry and is replaced by
\begin{eqnarray}
  %Vectors
  \partial_w V^i &=& - \delta^i{}_z \frac{V^w}{z}\,, \label{eq:CartVecFirst}\\
  \partial_i \partial_w V^j &=& \delta^j{}_z \left( - \frac{\partial_i V^w}{z}
                                 + \delta^z{}_i \frac{V^w}{z^2} \right) \,, \\
  \partial_w \partial_w V^w &=&
        \frac{\partial_z V^w}{z}- \frac{V^w}{z^2} \,. \label{eq:CartVecLast}
\end{eqnarray}
For rank two tensors,
Eq.~\eqref{eq:Cart2TensorZero} no longer holds. Instead, we have
\begin{eqnarray}
  \partial_w T_{\hat i \hat j} &=& 0 \, , \label{eq:CartTensFirst}\\
  \partial_w T_{iz} &=& - \frac{1}{z} T_{iw} - \delta_{zi} \frac{T_{zw}}{z} \, , \\
  \partial_w T_{ww} &=& 2\frac{T_{zw}}{z}\,,\\
  \partial_i \partial_w T_{\hat i \hat j} &=& 0 \, , \\
  \partial_i \partial_w T_{jz} &=& - \frac{\partial_i T_{jw} + \delta_{zj} \partial_i T_{wz}}{z}
                               + \delta_{iz} \frac{T_{jw} + \delta_{zj} T_{zw}}{z^2} \, , \\
  \partial_i \partial_w T_{ww} &=&   2\frac{\partial_i T_{zw}}{z}
                               -2 \delta_{iz} \frac{T_{zw}}{z^2} \,,\\
  \partial_w \partial_w T_{iw} &=& \frac{\partial_z T_{iw}}{z}
        -  \frac{T_{iw}+3\delta_{iz}T_{zw}}{z^2} \,. \label{eq:CartTensLast}
\end{eqnarray}
As for the case of $SO(D-d)$ symmetry the above expressions need to be regularized at $z=0$.
For Eqs. (\ref{eq:CartVecFirst}-\ref{eq:CartVecLast}), we note that $SO(2)$ symmetry implies that
vector components $V^w$ are odd functions of $z$ on the $w=0$ hyperplane. Therefore, the regularization of vector
components follows the procedure in Eqs. \eqref{eq:regPoint1} and \eqref{eq:regPoint4}.

For the regularization of Eqs. (\ref{eq:CartTensFirst}-\ref{eq:CartTensLast}), note that components of type $T_{\hat i w}$ behave like
vector components, that is they are odd functions of $z$. The component $T_{zw}$, on the other hand, has to vanish
at $z=0$ and must be an even function of $z$.
The latter can be seen by contracting $T_{\mu \nu}$ with two vectors pointing in the $w$ and $z$ direction respectively.
The result must be a scalar which satisfies the symmetry and is therefore even.
Together with the fact that $z$ and $w$ components of vectors are odd this then
implies that $T_{zw}$ is even.

Combining these relations with those previously discussed in
\ref{app:regularization}, we obtain the following regularized
terms specific to the case of $SO(2)$ symmetry.
\begin{eqnarray}
  \frac{V^w}{z} &\overset{*}{=}& \partial_z V^w\,,\\[10pt]
  -\frac{\partial_i V^w}{z} + \delta^z{}_i \frac{V^w}{z^2}
        &\overset{*}{=}& \left\{
    \begin{array}{ll}
      -\partial_i \partial_z V^w~~~~~&
        \text{if }i=\hat{i}\\[10pt]
      0 & \text{if }i = z
    \end{array}
    \right.
    \,, \\[10pt]
  -\frac{T_{iw}}{z} - \delta_{zi} \frac{T_{zw}}{z}
        &\overset{*}{=}& \left\{
    \begin{array}{ll}
      -\partial_z T_{iw}~~~~~&
        \text{if }i=\hat{i}\\[10pt]
      0 & \text{if }i = z
    \end{array}
    \right.
    \,, \\[10pt]
   -\frac{\partial_i T_{jw}+\delta_{jz}\partial_i T_{wz}}{z}
        + \delta_{iz}\frac{T_{jw}+\delta_{zj}T_{zw}}{z^2}
        &\overset{*}{=}& \left\{
    \begin{array}{ll}
      -\partial_i \partial_z T_{jw}~~~~~&
        \text{if }(i,j)=(\hat{i},\hat{j})\\[10pt]
      0 & \text{if }(i,j) = (z,\hat{j})\text{ or }(\hat{i},z) \\[10pt]
      -\partial_z \partial_z T_{wz} & \text{if }(i,j) = (z,z)
    \end{array}
    \right.
    \,, \nonumber\\
    && \\[10pt]
  2\frac{\partial_i T_{zw}}{z}-2\delta_{iz} \frac{T_{zw}}{z^2}
        &\overset{*}{=}& \left\{
    \begin{array}{ll}
        0 &
        \text{if }i = \hat{i}\\[10pt]
        \partial_z \partial_z T_{zw} ~~~~~& \text{if }i = z
    \end{array}
    \right.
    \,, \\[10pt]
  \frac{\partial_z T_{iw}}{z}-\frac{T_{iw}+3\delta_{iz}T_{zw}}{z^2}
        &\overset{*}{=}& \left\{
    \begin{array}{ll}
        0 &
        \text{if }i = \hat{i}\\[10pt]
        -\partial_z \partial_z T_{zw} ~~~~~& \text{if }i = z
    \end{array}
    \right.
    \,.
\end{eqnarray}
Finally, we list for completeness
the regularization of Eqs.~(\ref{eq:CartScalar2}),
(\ref{eq:CartVector1}), (\ref{eq:CartVector2}),
(\ref{eq:CartTensor2}), (\ref{eq:CartTensor3})
and (\ref{eq:CartTensor4}) expressed here in terms of generic
vector and tensor fields rather than the BSSN variables,
\begin{eqnarray}
  \frac{\partial_z \psi}{z} &\overset{*}{=}& \partial_z \partial_z \psi\,,
        \\
  \frac{V^z}{z} &\overset{*}{=}& \partial_z V^z\,,
        \\
  \frac{\partial_i V^z}{z} - \delta^z{}_i \frac{V^z}{z^2}
        &\overset{*}{=}& \left\{
        \begin{array}{ll}
          \partial_i \partial_z V^z ~~~~~&\text{if }i = \hat{i} \\[10pt]
          0 & \text{if }i = z
        \end{array}
        \right.
        \,,
        \\[10pt]
  \frac{\partial_z V^i}{z} - \delta^i{}_z \frac{V^z}{z^2}
        &\overset{*}{=}& \left\{
        \begin{array}{ll}
          \partial_z \partial_z V^i ~~~~~&\text{if }i = \hat{i} \\[10pt]
          0 & \text{if }i = z
        \end{array}
        \right.
        \,,
        \\[10pt]
  \frac{T_{zz}-T_{ww}}{z^2} &\overset{*}{=}& \frac{1}{2}
        \partial_z \partial_z (T_{zz} - T_{ww})\,,
%        \\
\end{eqnarray}
\begin{eqnarray}
  \frac{\partial_z T_{ww}}{z} &\overset{*}{=}&
        \partial_z \partial_z T_{ww}\,,
        \\
  \frac{T_{iz} - \delta_{iz} T_{ww}}{z}
        &\overset{*}{=}& \left\{
        \begin{array}{ll}
          \partial_z T_{iz}~~~~~&\text{if }i = \hat{i} \\[10pt]
          0 & \text{if }i = z
        \end{array}
        \right.
        \,, \\[10pt]
  \frac{\partial_i T_{jz} -\delta_{jz}\partial_i T_{ww}}{z}
        -\delta_{iz} \frac{T_{jz}-\delta_{jz}T_{ww}}{z^2}
        &\overset{*}{=}& \left\{
        \begin{array}{ll}
          \partial_i \partial_z T_{jz} ~~~~~&\text{if }(i,j)=(\hat{i},\hat{j})
          \\[10pt]
          0 & \text{if }(i,j)=(\hat{i},z)\text{ or }(z,\hat{j}) \\[10pt]
          \frac{\partial_z \partial_z (T_{zz}-T_{ww})}{2}
          & \text{if }(i,j)=(z,z)
        \end{array}
        \right.
        \,, \nonumber\\
        && \\[10pt]
  \frac{\partial_z T_{ij}}{z}
        - \frac{\delta_{iz}T_{jz} + \delta_{jz}T_{iz}
                -2\delta_{iz}\delta_{jz} T_{ww}}{z^2}
        &\overset{*}{=}& \left\{
        \begin{array}{ll}
          \partial_z \partial_z T_{ij} ~~~~~&\text{if }(i,j)=(\hat{i},\hat{j})
          \\[10pt]
          0 & \text{if }(i,j)=(\hat{i},z)\text{ or }(z,\hat{j})
          \\[10pt]
          \partial_z \partial_z T_{ww} &\text{if }(i,j)=(z,z)
        \end{array}
        \right.
        \,. \nonumber\\
        &&
\end{eqnarray}
%

%=============================================================================

\bibliographystyle{ws-ijmpd}
%\bibliography{newuli}

\begin{thebibliography}{10}

\bibitem{Pretorius:2004jg}
F.~Pretorius, {\em Class. Quantum Grav.} {\bf 22}  (2005) 425, gr-qc/0407110.

\bibitem{Abbott:2016blz}
B.~â. Abbott {\em et~al.}, {\em Phys. Rev. Lett.} {\bf 116}  (2016)   061102,
  arXiv:1602.03837 [gr-qc].

\bibitem{Read:2013zra}
J.~S. Read {\em et~al.}, {\em Phys. Rev. D} {\bf 88}  (2013)   044042,
  arXiv:1306.4065 [gr-qc].

\bibitem{Hinder:2013oqa}
I.~Hinder {\em et~al.}, {\em Class. Quant. Grav.} {\bf 31}  (2014)   025012,
  arXiv:1307.5307 [gr-qc].

\bibitem{Pretorius:2005gq}
F.~Pretorius, {\em Phys. Rev. Lett.} {\bf 95}  (2005)   121101, gr-qc/0507014.

\bibitem{Baker:2005vv}
J.~G. Baker, J.~Centrella, D.-I. Choi, M.~Koppitz and J.~van Meter, {\em Phys.
  Rev. Lett.} {\bf 96}  (2006)   111102, gr-qc/0511103.

\bibitem{Campanelli:2005dd}
M.~Campanelli, C.~O. Lousto, P.~Marronetti and Y.~Zlochower, {\em Phys. Rev.
  Lett.} {\bf 96}  (2006)   111101, gr-qc/0511048.

\bibitem{Cardoso:2014uka}
V.~Cardoso, L.~Gualtieri, C.~Herdeiro and U.~Sperhake, {\em Living Rev.
  Relativity} {\bf 18}  (2015)  ~1, arXiv:1409.0014 [gr-qc].

\bibitem{Emparan:2008eg}
R.~Emparan and H.~S. Reall, {\em {Living Reviews in Relativity}} {\bf 11}
  (2008) {http://www.livingreviews.org/lrr-2008-6}.

\bibitem{Myers:1986un}
R.~C. Myers and M.~J. Perry, {\em Annals Phys.} {\bf 172}  (1986)   304.

\bibitem{Shibata:2009ad}
M.~Shibata and H.~Yoshino, {\em Phys. Rev. D} {\bf 81}  (2010)   021501,
  arXiv:0912.3606 [gr-qc].

\bibitem{Shibata:2010wz}
M.~Shibata and H.~Yoshino, {\em Phys. Rev. D} {\bf 81}  (2010)   104035,
  arXiv:1004.4970 [gr-qc].

\bibitem{Gregory:1993vy}
R.~Gregory and R.~Laflamme, {\em Phys. Rev. Lett.} {\bf 70}  (1993) 2837,
  hep-th/9301052.

\bibitem{Lehner:2011wc}
L.~Lehner and F.~Pretorius  (2011) arXiv:1106.5184 [gr-qc].

\bibitem{Figueras:2015hkb}
P.~Figueras, M.~Kunesch and S.~Tunyasuvunakool, {\em Phys. Rev. Lett.} {\bf
  116}  (2016)   071102, arXiv:1512.04532 [hep-th].

\bibitem{Santos:2015iua}
J.~E. Santos and B.~Way, {\em Phys. Rev. Lett.} {\bf 114}  (2015)   221101,
  arXiv:1503.00721 [hep-th].

\bibitem{Bantilan:2012vu}
H.~Bantilan, F.~Pretorius and S.~S. Gubser, {\em Phys. Rev. D} {\bf 85}  (2012)
    084038, arXiv:1201.2132 [hep-th].

\bibitem{Chesler:2013lia}
P.~M. Chesler and L.~G. Yaffe, {\em JHEP} {\bf 1407}  (2014)   086,
  arXiv:1309.1439 [hep-th].

\bibitem{Bantilan:2014sra}
H.~Bantilan and P.~Romatschke, {\em Phys. Rev. Lett.} {\bf 114}  (2015)
  081601, arXiv:1410.4799 [hep-th].

\bibitem{Gubser:2014qua}
S.~S. Gubser and W.~van~der Schee, {\em JHEP} {\bf 01}  (2015)   028,
  arXiv:1410.7408 [hep-th].

\bibitem{Chesler:2015bba}
P.~M. Chesler, {\em Phys. Rev. Lett.} {\bf 115}  (2015)   241602,
  arXiv:1506.02209 [hep-th].

\bibitem{Buchel:2015saa}
A.~Buchel, M.~P. Heller and R.~C. Myers, {\em Phys. Rev. Lett.} {\bf 114}
  (2015)   251601, arXiv:1503.07114 [hep-th].

\bibitem{Antoniadis:1990ew}
I.~Antoniadis, {\em Phys. Lett. B} {\bf 246}  (1990) 377.

\bibitem{Antoniadis:1998ig}
I.~Antoniadis, N.~Arkani-Hamed, S.~Dimopoulos and G.~R. Dvali, {\em Phys. Lett.
  B} {\bf 436}  (1998) 257, hep-ph/9804398.

\bibitem{ArkaniHamed:1998rs}
N.~Arkani-Hamed, S.~Dimopoulos and G.~R. Dvali, {\em Phys. Lett. B} {\bf 429}
  (1998) 263, hep-ph/9803315.

\bibitem{Randall:1999ee}
L.~Randall and R.~Sundrum, {\em Phys. Rev. Lett.} {\bf 83}  (1999) 3370,
  hep-ph/9905221.

\bibitem{Randall:1999vf}
L.~Randall and R.~Sundrum, {\em Phys. Rev. Lett.} {\bf 83}  (1999) 4690,
  hep-th/9906064.

\bibitem{Banks:1999gd}
T.~Banks and W.~Fischler  (1999) hep-th/9906038.

\bibitem{Dimopoulos:2001hw}
S.~Dimopoulos and G.~Landsberg, {\em Phys. Rev. Lett.} {\bf 87}  (2001)
  161602, hep-th/0106295.

\bibitem{Giddings:2001bu}
S.~B. Giddings and S.~Thomas, {\em Phys. Rev. D} {\bf 65}  (2002)   056010,
  hep-ph/0106219.

\bibitem{Sperhake:2008ga}
U.~Sperhake, V.~Cardoso, F.~Pretorius, E.~Berti and J.~A. Gonz{\'a}lez, {\em
  Phys. Rev. Lett.} {\bf 101}  (2008)   161101, arXiv:0806.1738 [gr-qc].

\bibitem{Shibata:2008rq}
M.~Shibata, H.~Okawa and T.~Yamamoto, {\em Phys. Rev. D} {\bf 78}  (2008)
  101501(R), arXiv:0810.4735 [gr-qc].

\bibitem{Choptuik:2009ww}
M.~W. Choptuik and F.~Pretorius, {\em Phys. Rev. Lett.} {\bf 104}  (2010)
  111101, arXiv:0908.1780 [gr-qc].

\bibitem{Sperhake:2009jz}
U.~Sperhake, V.~Cardoso, F.~Pretorius, E.~Berti, T.~Hinderer and N.~Yunes, {\em
  Phys. Rev. Lett.} {\bf 103}  (2009)   131102, arXiv:0907.1252 [gr-qc].

\bibitem{East:2012mb}
W.~E. East and F.~Pretorius, {\em Phys. Rev. Lett.} {\bf 110}  (2013)   101101,
  arXiv:1210.0443 [gr-qc].

\bibitem{Rezzolla:2012nr}
L.~Rezzolla and K.~Takami, {\em Class. Quant. Grav.} {\bf 30}  (2013)   012001,
  arXiv:1209.6138 [gr-qc].

\bibitem{Sperhake:2012me}
U.~Sperhake, E.~Berti, V.~Cardoso and F.~Pretorius, {\em Phys. Rev. Lett.} {\bf
  111}  (2013)   041101, arXiv:1211.6114 [gr-qc].

\bibitem{Healy:2015mla}
J.~Healy, I.~Ruchlin and C.~O. Lousto  (2015) arXiv:1506.06153 [gr-qc].

\bibitem{Sperhake:2015siy}
U.~Sperhake, E.~Berti, V.~Cardoso and F.~Pretorius  (2015) arXiv:1511.08209
  [gr-qc].

\bibitem{Witek:2010xi}
H.~Witek, M.~Zilh{\~a}o, L.~Gualtieri, V.~Cardoso, C.~Herdeiro, A.~Nerozzi and
  U.~Sperhake, {\em Phys. Rev. D} {\bf 82}  (2010)   104014, arXiv:1006.3081
  [gr-qc].

\bibitem{Witek:2010az}
H.~Witek, V.~Cardoso, L.~Gualtieri, C.~Herdeiro, U.~Sperhake and M.~Zilh{\~a}o,
  {\em Phys. Rev. D} {\bf 83}  (2011)   044017, arXiv:1011.0742 [gr-qc].

\bibitem{Okawa:2011fv}
H.~Okawa, K.-i. Nakao and M.~Shibata, {\em Phys. Rev. D} {\bf 83}  (2011)
  121501, arXiv:1105.3331 [gr-qc].

\bibitem{Witek:2014mha}
H.~Witek, H.~Okawa, V.~Cardoso, L.~Gualtieri, C.~Herdeiro, M.~Shibata,
  U.~Sperhake and M.~Zilh{\~a}o, {\em Phys. Rev. D} {\bf 90}  (2014)   084014,
  arXiv:1406.2703 [gr-qc].

\bibitem{Geroch:1970nt}
R.~Geroch, {\em J. Math. Phys.} {\bf 12}  (1970) 918.

\bibitem{Zilhao:2010sr}
M.~Zilh{\~a}o, H.~Witek, U.~Sperhake, V.~Cardoso, L.~Gualtieri, C.~Herdeiro and
  A.~Nerozzi, {\em Phys. Rev. D} {\bf 81}  (2010)   084052, arXiv:1001.2302
  [gr-qc].

\bibitem{Zilhao:2013gu}
M.~Zilh{\~a}o, {New frontiers in Numerical Relativity}, PhD thesis, University
  of Porto  (2012).
\newblock arXiv:1301.1509 [gr-qc].

\bibitem{Alcubierre:1999ab}
M.~Alcubierre, S.~Brandt, B.~Br{\"u}gmann, D.~Holz, E.~Seidel, R.~Takahashi and
  J.~Thornburg, {\em Int. J. Mod. Phys. D} {\bf 10}  (2001) 273, gr-qc/9908012.

\bibitem{Yoshino:2011zz}
H.~Yoshino and M.~Shibata, {\em Prog.Theor.Phys.Suppl.} {\bf 189}  (2011) 269.

\bibitem{Yoshino:2011zza}
H.~Yoshino and M.~Shibata, {\em Prog.Theor.Phys.Suppl.} {\bf 190}  (2011) 282.

\bibitem{Shibata:1995we}
M.~Shibata and T.~Nakamura, {\em Phys. Rev. D} {\bf 52}  (1995) 5428.

\bibitem{Baumgarte:1998te}
T.~W. Baumgarte and S.~L. Shapiro, {\em Phys. Rev. D} {\bf 59}  (1998)
  024007, gr-qc/9810065.

\bibitem{Arnowitt:1962hi}
R.~Arnowitt, S.~Deser and C.~W. Misner, { The dynamics of general relativity},
  in {\em Gravitation an introduction to current research\/},  ed. L.~Witten
  (John Wiley, New York, 1962), pp. 227--265.
\newblock gr-qc/0405109.

\bibitem{York1979}
J.~W. York, Jr., { Kinematics and dynamics of general relativity}, in {\em
  Sources of {G}ravitational {R}adiation\/},  ed. L.~Smarr (Cambridge
  University Press, Cambridge, 1979), pp. 83--126.

\bibitem{Gourgoulhon:2007ue}
E.~Gourgoulhon  (2007) gr-qc/0703035.

\bibitem{Marronetti:2007wz}
P.~Marronetti, W.~Tichy, B.~Br{\"u}gmann, J.~A. Gonz{\'a}lez and U.~Sperhake,
  {\em Phys. Rev. D} {\bf 77}  (2008)   064010, arXiv:0709.2160 [gr-qc].

\bibitem{Alcubierre:2002kk}
M.~Alcubierre, B.~Br{\"u}gmann, P.~Diener, M.~Koppitz, D.~Pollney, E.~Seidel
  and R.~Takahashi, {\em Phys. Rev. D} {\bf 67}  (2003)   084023,
  gr-qc/0206072.

\bibitem{Yo:2002bm}
H.-J. Yo, T.~W. Baumgarte and S.~L. Shapiro, {\em Phys. Rev. D} {\bf 66}
  (2002)   084026, gr-qc/0209066.

\bibitem{Emparan:2001wn}
R.~Emparan and H.~S. Reall, {\em Phys. Rev. Lett.} {\bf 88}  (2002)   101101,
  hep-th/0110260.

\bibitem{Figueras:2016}
P.~Figueras, M.~Kunesch and S.~Tunyasuvunakool  (in preparation).

\bibitem{Tangherlini:1963bw}
F.~Tangherlini, {\em Nuovo Cim.} {\bf 27}  (1963) 636.

\bibitem{Brill:1963yv}
D.~R. Brill and R.~W. Lindquist, {\em Phys. Rev.} {\bf 131}  (1963) 471.

\bibitem{Sperhake:2006cy}
U.~Sperhake, {\em Phys. Rev. D} {\bf 76}  (2007)   104015, gr-qc/0606079.

\bibitem{Sperhake:2007gu}
U.~Sperhake, E.~Berti, V.~Cardoso, J.~A. Gonz{\'a}lez, B.~Br{\"u}gmann and
  M.~Ansorg, {\em Phys. Rev. D} {\bf 78}  (2008)   064069, arXiv:0710.3823
  [gr-qc].

\bibitem{Cactusweb}
{Cactus Computational Toolkit homepage:} {\tt http://www.cactuscode.org/}.

\bibitem{Allen:1999}
{Allen, G. and Goodale, T. and Mass\'o, J. and Seidel, E.}, { {The Cactus
  Computational Toolkit and Using Distributed Computing to Collide Neutron
  Stars}}, in {\em {Proceedings of Eighth IEEE International Symposium on High
  Performance Distributed Computing, HPDC-8, Redondo Beach, 1999}\/},  ({IEEE
  Press}, {}, 1999).

\bibitem{Carpetweb}
{Carpet Code homepage}: {\tt http://www.carpetcode.org/}.

\bibitem{Schnetter:2003rb}
E.~Schnetter, S.~H. Hawley and I.~Hawke, {\em Class. Quant. Grav.} {\bf 21}
  (2004) 1465, gr-qc/0310042.

\bibitem{vanMeter:2006vi}
J.~R. van Meter, J.~G. Baker, M.~Koppitz and D.-I. Choi, {\em Phys. Rev. D}
  {\bf 73}  (2006)   124011, gr-qc/0605030.

\bibitem{Garfinkle:2001ni}
D.~Garfinkle, {\em Phys. Rev. D} {\bf 65}  (2002)   044029, gr-qc/0110013.

\bibitem{Alic:2011gg}
D.~Alic, C.~Bona-Casas, C.~Bona, L.~Rezzolla and C.~Palenzuela, {\em Phys. Rev.
  D} {\bf 85}  (2012)   064040, arXiv:1106.2254 [gr-qc].

\bibitem{Weyhausen:2011cg}
A.~Weyhausen, S.~Bernuzzi and D.~Hilditch, {\em Phys. Rev. D} {\bf 85}  (2012)
   024038, arXiv:1107.5539 [gr-qc].

\end{thebibliography}

\end{document}